\documentstyle[12pt,epsfig]{article}
\begin{document}

\begin{titlepage}
\rightline{June 2015}
\vskip 2cm
\centerline{\Large \bf
Dissipative dark matter  
}
\vskip 0.34cm
\centerline{\Large \bf
and the rotation curves of dwarf galaxies}

\vskip 1.7cm
\centerline{R. Foot\footnote{
E-mail address: rfoot@unimelb.edu.au}}

\vskip 0.7cm
\centerline{\it ARC Centre of Excellence for Particle Physics at the Terascale,}
\centerline{\it School of Physics, University of Melbourne,}
\centerline{\it Victoria 3010 Australia}
\vskip 3.8cm
\noindent
There is ample evidence from rotation curves that dark matter halos around disk galaxies have nontrivial dynamics.
Of particular significance are: a) the cored dark matter profile of disk galaxies, b) correlations of
the shape of rotation curves with baryonic properties, and c)
Tully-Fisher relations.
Dark matter halos around disk galaxies may have nontrivial dynamics if dark matter
is strongly self interacting and dissipative. 
Multicomponent hidden sector dark matter featuring a massless `dark photon' (from an unbroken dark $U(1)$ gauge
interaction) which kinetically mixes with the ordinary photon provides a concrete 
example of such dark matter. The kinetic mixing interaction facilitates halo heating by enabling
ordinary supernovae to be a source of these `dark photons'. 
Dark matter halos can expand and contract
in response to the heating and cooling processes, but for a sufficiently isolated halo could
have evolved to a steady state or `equilibrium' 
configuration where heating and cooling rates locally balance. 
This dynamics allows the dark matter density profile to be related to the distribution of ordinary supernovae
in the disk of a given galaxy. 
In a previous paper a simple and predictive formula was derived  encoding this relation.
Here we improve on previous work by modelling the supernovae distribution via the measured UV and $H\alpha$ fluxes,
and
compare the resulting dark matter halo profiles with the rotation curve data for each dwarf galaxy in the LITTLE THINGS sample.
The dissipative dark matter concept is further developed and some conclusions drawn.

\end{titlepage}

\section{Introduction}

%\vskip 0.3cm

Large scale structure (LSS) and measurements of the  cosmic microwave background (CMB) provide strong evidence for the 
existence of nonbaryonic dark
matter in the Universe, e.g. \cite{planck,SPT,WMAP,LSS6,LSS5,LSS4,LSS3,LSS1,LSS2}.  
Dark matter, whether collisionless or collisional, provides a compelling explanation
for the measured anisotropies of the CMB and the observed large scale structure.
On much smaller scales,
measurements of the rotation curves of disk galaxies have also yielded very convincing evidence
for the existence of nonbaryonic dark matter. Indeed, galaxy rotation curves are
(typically) observed to be flat at the observed edge of the galaxy, in sharp contrast to the
expected Keplerian decline \cite{rubinhis,roberts,rubinhis2,bosma} (see also \cite{rubin} and references therein). 

This asymptotic flatness is certainly an intriguing feature, yet rotation curves 
provide much more information.
Detailed studies, especially those involving low surface brightness galaxies, e.g.  \cite{LSB,LSB2,LSB3} and 
gas rich dwarfs, e.g. \cite{thingsdwarfs,oh,swaters}, have found that the
dark matter profile is generally cored (see also the review \cite{review}) and that  
the shape of rotation curves are correlated with baryonic properties. There are also local such correlations observed in
specific galaxies, such as the rotation curve `wiggle' in NGC1560 \cite{bor,gen}. Similar conclusions have 
also been reached in high surface brightness spiral galaxies \cite{sal,sal2,sal1}. 
See also \cite{kent,sanx,ugc,donato1,stacy,lelli} and references therein 
for further relevant discussions.
Such features, and others (e.g.  Tully-Fisher relations \cite{tf,btf})
suggests to this author that dark matter self interactions are likely to be important on small scales, cf. \cite{Moore,Spergel,kapx}.

In this paper we consider a particular class of collisional dark matter, dissipative dark matter.
By this we mean
dark matter with strong self interactions which include dissipative
particle processes involving the emission of a massless (or very light) dark bosonic particle. 
That is, dark matter is envisaged to have broadly similar
particle properties to ordinary matter. 
Multicomponent hidden sector dark matter featuring a massless `dark photon' (from an unbroken dark $U(1)$ gauge
interaction) which kinetically mixes with the ordinary photon provides a concrete 
example of such dark matter \cite{flv,footvolkas,footrevold,foot4}
(see also the review \cite{footreview} for a more detailed bibliography).
\footnote{More generally, hidden sector dark matter models with an unbroken dark $U(1)$ gauge interaction have been quite widely discussed 
in the literature, e.g. \cite{hall,hid1,hidb,feng,cline,fan,petraki,estonia,pet2}.}
The kinetic mixing interaction \cite{foothe,holdom,flv},
\begin{eqnarray}
{\cal L}_{int} = - \frac{\epsilon}{2}F^{\mu \nu} F_{\mu \nu}^{'} \ ,
\label{1x}
\end{eqnarray}
facilitates halo heating by enabling
ordinary supernovae to be a source of these `dark photons' \cite{raf,footsilold}. These dark photons can transport a large fraction
of a supernova's core collapse energy to the halo (potentially up to $\sim 10^{53}$ ergs per supernova for kinetic mixing of
strength $\epsilon \sim 10^{-9}$). 
Dark matter halos can thereby expand and contract
in response to these heating and cooling processes, but at the current epoch 
could (typically) have evolved to a steady state or `equilibrium' 
configuration where heating and cooling rates locally balance. 

The resulting dynamics allows the dark matter density profile to be related to the distribution of ordinary supernovae
in the disk of a given galaxy. 
In a previous paper \cite{foot5} a simple formula was derived  encoding this relation 
which, it was argued, should approximately represent dissipative dynamics independently of the details
of the particular dissipative particle physics model.
This formula is highly predictive as
all of the model dependence, fundamental physics etc., was able to be condensed into a single parameter, $\lambda$.
It was shown that this approach  could potentially explain the apparent correlations between 
the dark matter and ordinary matter distributions in galaxies.
The purpose of the present paper is to further explore this
predicted halo profile of dissipative dark matter, examining all 26 dwarf galaxies in the LITTLE THINGS sample \cite{oh}.
In doing so, we aim to improve on previous work by modelling the supernovae distribution via the measured UV and $H\alpha$ fluxes.
This is a more direct measure of the supernovae distribution than that considered previously in \cite{foot5} where a
Kennicutt-Schmidt type relation \cite{Ken,Sch} was used to model the supernovae distribution in terms of the baryonic gas density. 
We also briefly consider THINGS spirals \cite{things}. Although spirals are generally not as useful in testing halo profiles due
to the much larger baryonic contribution, they are important, especially for exploring the scaling behaviour of $\lambda$.

\section{Halo profile from dissipative dark matter}

It has been known for some time that 
dissipative dark matter models
can explain dark matter phenomena on large scales: LSS and CMB anisotropies \cite{foot13,canada,berezhiani}.
These types of models are also
consistent with other cosmological probes, such as $\delta N_{eff} (BBN)$ and $\delta N_{eff}(CMB)$ \cite{cfother,foot4}.
The small scale structure of dark matter, especially the dark matter distribution around galaxies,
has been pursued in \cite{footvolkas,foot1,foot2,foot3,footsil,foot4,footreview,foot5}. 

In this picture the halo of disk galaxies is in the form of a strongly self interacting
gas of particles. In the model of \cite{foot4} this gas is a plasma composed of dark electrons ($F_1$)
and dark ions ($F_2$) interacting via massless dark photons, $\gamma_D$.
Such a halo is dynamical and can be modelled as a fluid governed by Euler's equations, with both
heating and cooling processes.
As in the previous studies, we assume the existence of  
a kinetic  mixing interaction, Eq.(\ref{1x}), 
of strength $\epsilon \sim 10^{-9}$.
This small interaction results in a substantial halo heat source, with up to around half
of the total core-collapse energy of ordinary supernovae converted into dark photons \cite{raf,footsilold,foot1}.
These dark photons propagate out into the halo where they can eventually be
absorbed there via some interaction process, with dark photoionization a 
likely suspect in the specific models studied \cite{footvolkas,foot1,foot2,foot3,foot4}. 
It follows that
the heating rate at a particular point, $P$, in the halo is proportional to 
the product of the dark matter density and dark photon energy flux at that point: $\Gamma_{heat}({\bf r})
\propto n({\bf r})  F_{\gamma_D} (\bf{r})$.
The halo is dissipative, and cools via dark bremsstrahlung (and potentially other processes),
which means that the cooling rate at the point, $P$, is proportional to the square of dark matter density:
$\Gamma_{cool}({\bf r})  \propto n({\bf r})^2$.
[The proportionality coefficients depend on the details of the dissipative particle physics, and do not 
need to be specified for the purposes of model independent considerations.]

Given sufficient time, a dark matter halo can evolve to a steady-state configuration where
the heating and cooling rates balance at every location, $\Gamma_{heat}({\bf r}) = \Gamma_{cool}({\bf r})$, so that: 
\begin{eqnarray}
n({\bf r}) \propto F_{\gamma_D} ({\bf r}).
\label{sunx}
\end{eqnarray}
The timescale of this halo evolution, $\tau$, is presumed to be much less than the current age of the Universe.
%which is easily satisfied in specific models, e.g. \cite{foot4}. 
A rough estimate for this timescale is the time for which dissipative interactions would dissipate all
the halo's energy in the absence of heating, i.e.  $\tau \sim (3/2)n(r)T/\Gamma_{cool}$ (here
$T$ is the halo temperature).
The precise value of $\tau$ is of course model
dependent, but expected to be less than around a Gyr in the specific models studied \cite{foot4,footreview}.

As discussed earlier,
the flux of dark photons which govern this dynamics is expected to originate within 
and around core-collapse supernovae. Supernovae are the final evolutionary stages of
large ($M \stackrel{>}{\sim} 8 m_{\odot}$) stars which are located in the
galactic disk.
Using spherical coordinates and
setting the presumed thin disk at $\theta = \pi/2$ (i.e. the $z$ axis is normal to the plane of the disk) 
the energy flux of these dark photons at a point $P = (r,\theta,\phi)$ 
within an optically thin halo is given by:
\footnote{
Supernovae produce a flux of dark photons with an uncertain energy spectrum.
%Optically thin means that 
%the optical depth, $\tau$, along the dark photon's path between
%its production point and absorption point satisfies $\tau \stackrel{<}{\sim} 1$.
In general the optical depth is a frequency (energy) dependent quantity so that
in actuality the halo might only be optically
thin for a range of dark photon frequencies. However these optically thin dark photons 
can dominate the energy transport, making the optically thin approximation potentially valid even in  
this case. 
This conclusion is also supported by the numerical work of \cite{foot2,foot3} where the effects of
finite optical depth was considered where the uncertain dark photon supernova spectrum was modelled 
with a wide range of possible dark photon frequency distributions.}
\begin{eqnarray}
F_{\gamma_D} (r,\theta,\phi) \propto  
\int
d\widetilde{\phi} \int d\widetilde{r} \ \widetilde{r} 
\ \frac{\Sigma_{SN}
({\widetilde{r},\widetilde{\phi}})}{4\pi[r^2 + {\widetilde{r}}^2 - 2r\widetilde{r}  \sin\theta \cos
(\widetilde{\phi}-\phi)]}
\ .
\label{1}
\end{eqnarray}
Here, $\Sigma_{SN} (\widetilde{r},\widetilde{\phi})$
is the type II supernova rate per unit area in the disk.
Strictly, $\Sigma_{SN}$ is some appropriately weighted average over the timescale $\sim \tau$, expected to be many orders
of magnitude greater than the time between the discrete supernova events.
%Since supernovae are discrete events, this quantity represents an average over a reasonable time period (at least 10 million years).

The dynamically driven balancing of heating and cooling rates [$\Gamma_{heat}({\bf r}) = \Gamma_{cool} ({\bf r})$]
dictates the dark matter density, via Eq.(\ref{sunx}). That is,
\begin{eqnarray}
\rho(r,\theta,\phi) =  
\lambda \int
d\widetilde{\phi} \int d\widetilde{r} \ \widetilde{r} 
\ \frac{\Sigma_{SN}
({\widetilde{r},\widetilde{\phi}})} {4\pi[r^2 + {\widetilde{r}}^2 - 2r\widetilde{r}  \sin\theta \cos
(\widetilde{\phi}-\phi)]}
\ .
\label{3z}
\end{eqnarray}
The proportionality coefficient, $\lambda$, depends on the dark photoionization cross section, 
supernovae dark photon energy spectrum, and halo properties: ionization state, composition and temperature.
Evidently, there is a significant parameter degeneracy which potentially makes this kind of dark matter model
extremely predictive as far as galaxy dynamics is concerned, despite the nontrivial nature of dissipative dark matter.

A critical assumption 
in deriving Eq.(\ref{3z}) is that the dark halo is able to dynamically evolve until it reaches a steady-state configuration.
Naturally, this could only occur
if the galaxy is sufficiently isolated and the heat source (supernova rate) is sufficiently stable.
Environmental effects, such
as mergers, perturbations from nearby galaxies etc. could hinder an actual
galaxy from attaining this `equilibrium' configuration. The degree to which a particular galaxy
is perturbed away from equilibrium would depend on the size of the perturbation
as well as when it occurred.  Once a perturbation ceases, the galaxy can re-equilibrate on the timescale 
$\sim \tau$, which as mentioned above is expected to be less than a Gyr.
An important class of `perturbed' galaxies are those currently in a starburst phase. Starburst galaxies exhibit a rapidly varying
star formation rate over the last few hundred million years. 
Given that ordinary supernovae are the presumed
source of the dark photons which heat the halo,    
this then implies a rapidly changing halo heating rate [$\Gamma_{heat}({\bf r},t)$].
Such a rapidly changing halo heat source   
can, of course, potentially perturb the halo away from the steady-state configuration. Thus, the density profile
of the dark matter halo of starburst galaxies can depart significantly from Eq.(\ref{3z}). 

%% paragraph about model indepdence here %%

Although this discussion has been couched in terms of a specific kind of dissipative 
dark matter model (where halo heating is supplied by massless dark photons produced via kinetic mixing induced
processes in ordinary supernovae) Eq.(\ref{3z}) can hold more generally.
The basic assumptions are only that dark matter is dissipative, behaves as a fluid which evolves to a steady-state configuration
where heating and cooling rates equilibrate (on a timescale $\tau \stackrel{<}{\sim} 1$ Gyr).  
It is easy therefore to envisage many other dissipative dark matter models leading to Eq.(\ref{3z}).
For example, models where heating is transported to the halo via massive dark photons are possible. Alternatively 
heating might be transported from ordinary supernovae to the halo via light scalar particles.  In these scenarios,
various considerations constrain the 
mass of the massive dark photon or dark
scalar to be around $\sim$ 10 keV. 
However, models with energy sources other than supernovae, e.g. ordinary main sequence stars, are unlikely
given the typically stringent constraints on exotic energy loss mechanisms for such objects.     

%% finish paragraph/

In general, the coefficient $\lambda$ in Eq.(\ref{3z})
will depend on the position, $r, \theta, \phi$ and can be model dependent.
%The size of this spatial variation is expected to depend on the details of the particular dissipative dark matter model.
%For example, one might be concerned with
Such spatial variation of $\lambda$ is due,
in part, to the dependence on
$\lambda$ on the halo temperature. 
Simplified calculations, though, within a given dissipative model \cite{foot2,footreview}
suggest that the spatial variation of the temperature is only important
in the central region of the galaxy and even in that region can (typically) be relatively modest for
small galaxies. 
%Of course, even for small galaxies, it is possible that the spatial dependence of $\lambda$ could be important 
%for particular galaxies.
%For instance, in regions where the halo temperature of a given galaxy happens to be close to an important ionization
%transition temperature; the details depending on the particular dissipative dark matter model.
For the purposes of the analysis given here we shall consider  
$\lambda$ in Eq.(\ref{3z}) to be spatially independent  
as a zeroth order approximation. 
Under this approximation the
spatial dependence of $\rho (r,\theta,\phi)$, Eq.(\ref{3z}), is determined solely by
that of the flux, $F_{\gamma_D}(r,\theta,\phi)$.

Eq.(\ref{3z}) is subject to another significant caveat. It provides the mass density only of the diffuse dark matter  
(plasma) component.
Dissipative dark matter models can also have dark matter in
the form of `dark stars', that is, stars  composed of dark matter particles. Naturally it is quite
challenging to figure out the proportion and distribution of such a component, and for the present
discussion it will be assumed negligible. 
\footnote{Note though that for elliptical galaxies (and possibly dwarf spheroidals) the situation is 
expected to be very different. For these galaxy types there is (currently) very low star formation rate  
and therefore heating of the dark halo from ordinary supernovae is not expected to be significant.
Any diffuse dissipative dark matter halo around these kinds of galaxies would therefore have collapsed into `dark stars' \cite{footreview,foot4}.}

With the above important qualifications, Eq.(\ref{3z}) provides a description of the dark matter density 
in a given galaxy. Disk galaxies though come in various sizes, how does $\lambda$ depend on the
galaxy scale? The physical quantity of most interest is again the halo temperature, $T$. Assuming that
the halo is isothermal and in approximate hydrostatic equilibrium, 
the temperature can be expressed in terms of the maximum value
of the rotational velocity: \cite{footvolkas,foot4,footreview}
\begin{eqnarray}
T = {1 \over 2} \bar m \left[v_{rot}^{max}\right]^2
\label{T}
\end{eqnarray}  
where $\bar m = \sum n_i m_i/\sum n_i$ is the mean mass of the particles making up the dark plasma.
The maximum value of the rotational velocity, $v_{rot}^{max}$, varies between around $20$ km/s for the smallest
dwarfs to around 300 km/s for the largest spirals.
If we assume that
the heating is due to photoionization of K-shell dark atomic states and that the binding energy of these 
states is larger than the halo temperature of the largest spirals,
then the heating rate is approximately independent of the halo's temperature.
The dark photons heating the halo encounter fully occupied K-shell states for all galaxy scales of interest.
Cooling, on the other hand, is expected to depend more strongly on the halo's temperature
since the bremsstrahlung cross section depends on temperature, leading to:
$\Gamma_{cool} \propto \sqrt{T} n({\bf r})^2$.
Thus, these simplified dissipative models suggest that $\lambda \propto 1/\sqrt{T}$, and via Eq.(\ref{T}) 
give $\lambda \propto 1/v_{rot}^{max}$. 
However more complicated behaviour is certainly possible, and would be expected for temperature regions where other
cooling processes become important, 
\footnote{
For an important class of dissipative models, including those studied in  
\cite{footvolkas,foot1,foot2,foot3,footreview,foot4},
dark atom recombination and line emission can potentially significantly contribute to cooling. 
These processes also feature $\Gamma_{cool} \propto n({\bf r})^2$
and thus Eq.(\ref{3z}) still holds,
but the proportionality constant can have a more complicated dependence on halo temperature.
If such processes are dynamically important they can modify the predicted $\lambda(T) \propto 1/\sqrt{T}$ 
scaling. }
the details will of course depend on the particle physics 
properties of the specific dissipative model.  
As will be discussed in more detail in section 3, this galaxy scaling relation
is related to the Tully-Fisher relation \cite{tf}. 
%We will examine the validity of this scaling relation
%for the LITTLE THINGS and THINGS samples at the end of section 3.

The halo dark matter contribution to the gravitational 
acceleration at a point in the plane of the disk
can be straightforwardly calculated given the density profile, Eq.(\ref{3z}), and Newton's law of gravity. 
Assuming an azimuthally symmetric disk, the
motion of the gas and stars in the disk is circular with speed $v_{halo}$ given by:
\begin{eqnarray}
%{v^2_{halo}  \over r} = G_N \int d\widetilde{\phi} \int d\cos\widetilde{\theta} 
%\int d\widetilde{r} \ \widetilde{r}^2 \ {\rho (\widetilde{r},\widetilde{\theta}) \cos\omega \over 
%r^2 + {\widetilde{r}}^2 - 2r \widetilde{r}  \sin\widetilde{\theta} \cos \widetilde{\phi}}
{v^2_{halo}  \over r} = G_N \int d\widetilde{\phi} \int d\cos\widetilde{\theta} 
\int d\widetilde{r} \ \widetilde{r}^2 \ {\rho (\widetilde{r},\widetilde{\theta}) \cos\omega \over 
d^2}
%r^2 + {\widetilde{r}}^2 - 2r \widetilde{r}  \sin\widetilde{\theta} \cos
%\widetilde{\phi}}
\ .
\label{4z}
\end{eqnarray}
Here, $d^2 \equiv r^2 + {\widetilde{r}}^2 - 2r \widetilde{r}  \sin\widetilde{\theta} \cos \widetilde{\phi}$,  
%\begin{eqnarray} 
$\cos\omega \equiv (r - \widetilde{r}\sin\widetilde{\theta}\cos\widetilde{\phi})/d$ and
$G_N$ is Newton's constant. 
%(\sqrt{r^2 + {\widetilde{r}}^2 - 2r \widetilde{r}  \sin\widetilde{\theta} \cos
%\widetilde{\phi}})$.
%\end{eqnarray}

In the previous paper \cite{foot5} the density profile, Eq.(\ref{3z}), was 
examined by first modelling the supernovae rate
with an exponential disk: $\Sigma_{SN} (r) = (R_{SN}/2\pi r_D^2) \ e^{-r/r_D}$ \cite{freeman}. 
This could only be a very crude approximation, since the exponential disk is an approximate measure of the 
stellar population, including middle aged and 
older stars, while the quantity we require is the recent (large) star formation rate.
Anyway, if $\Sigma_{SN}(r)$ is an exponential function of radius
then
Eq.(\ref{3z}) becomes roughly equivalent 
(as far as rotation curves are concerned) to the cored distribution:
$\rho_{ISO} (r) = \rho_0 r_0^2/(r^2 + r_0^2)$, with $r_0 \approx r_D$.
Such a constrained quasi-isothermal profile
is known to be phenomenologically successful in explaining the observed shapes of 
rotation curves, e.g. \cite{LSB,blok,blok2,salucci,donato1,oh}.
Dissipative dark matter therefore provides an underlying theoretical explanation 
for the successful quasi-isothermal profile.

The dark matter density profile of Eq.(\ref{3z}) is, in fact, further constrained.
At small radii, $r < r_D$, the dark matter density [Eq.(\ref{3z})]
can be related to the central surface density, $\Sigma_{SN} (0) \equiv R_{SN}/(2\pi r_D^2)$, via:
\begin{eqnarray}
\rho (r) = \frac{\lambda \Sigma_{SN} (0)}{2}\left[ log\left(\frac{r_D}{r}\right) \  +  \ {\rm constant} \right]
\ .
\end{eqnarray}
This implies a scaling relation 
connecting the inner circular velocity gradient with this central surface density:
$v \propto r \sqrt{\lambda \Sigma_{SN} (0)}$. 
To connect with measurable quantities, one could replace $\Sigma_{SN}(0)$ with the
central surface brightness in the UV band, $\Sigma_{UV}(0)$ and
neglect  the variation of $\lambda$ (which is anticipated 
to be fairly weak for dwarf irregular galaxies).
This leads to the
rough scaling relation: $v \propto r\sqrt{\Sigma_{UV} (0)}$ or
\begin{eqnarray}
log\left( \frac{dv}{dr}\right) = -0.2\mu_0 \  +  \ {\rm constant}  
\end{eqnarray}
where $\mu_0$ is the central surface brightness in magnitude units.
Such a scaling relation has in fact been observed to hold for spiral and irregular galaxies \cite{lelli}.

In addition to modelling $\Sigma_{SN}$ via an exponential disk, the previous paper \cite{foot5}
further considered
modelling $\Sigma_{SN}$ in terms of the baryonic gas density via a Kennicutt-Schmidt 
type relation \cite{Ken,Sch}.
This allowed a connection between the supernovae distribution with current local properties of a given galaxy.
%and would allow one to consider the azimuthally asymmetric case.
In the following we shall consider a more direct estimate of the supernovae distribution,
by modelling the supernovae distribution via the measured UV and $H\alpha$ fluxes.

\section{Dissipative dark matter versus dwarf galaxies}
\subsection{The LITTLE THINGS dwarfs}

In this paper we shall examine all 26 dwarf galaxies
comprising the LITTLE THINGS sample \cite{oh}.
An inspection of these 26 galaxies suggests a loose classification:
18 dwarfs 
which feature `classically' shaped rotation curves with a (typically) linear rise in $v_{rot}$ near $r=0$
smoothly transiting to a flat rotation curve at the greatest measured radii,
three dwarfs which show a classically shaped rotation curve for $r$ less than some radius, $R^*$, but show a downturn for $r > R^*$, and
four dwarfs which show a classically shaped
rotation curve for $r$ less than some radius, $R_1$, but feature a `hump' at $r > R_1$. In addition there is one dwarf (DDO47)
whose rotation curve is irregularly shaped.
In figure 1 we give an example for each of these galaxy types. 

The theoretically predicted rotation velocity, $v_{rot}$, is 
the sum of the various contributions added in quadrature:
\begin{eqnarray}
v_{rot} = \sqrt{v_{halo}^2 + v_{gas}^2 + v_{stars}^2} 
\ .
\label{rotx}
\end{eqnarray}
Here, $v_{gas}$ and $v_{stars}$ are the baryonic gas and stellar contributions
while $v_{halo}$ is the dark matter contribution,   
Eq.(\ref{4z}). The latter is given in terms of the density, Eq.(\ref{3z}), which can be evaluated if the
supernovae distribution of the galaxy is known.
To proceed we
therefore need to model the supernovae rate, $\Sigma_{SN} (r)$.

\vskip 1.0cm
\hspace{0.11cm}
%\centerline{\epsfig{file=fig1.eps,angle=0,width=11.0cm}}
\leftline{\epsfig{file=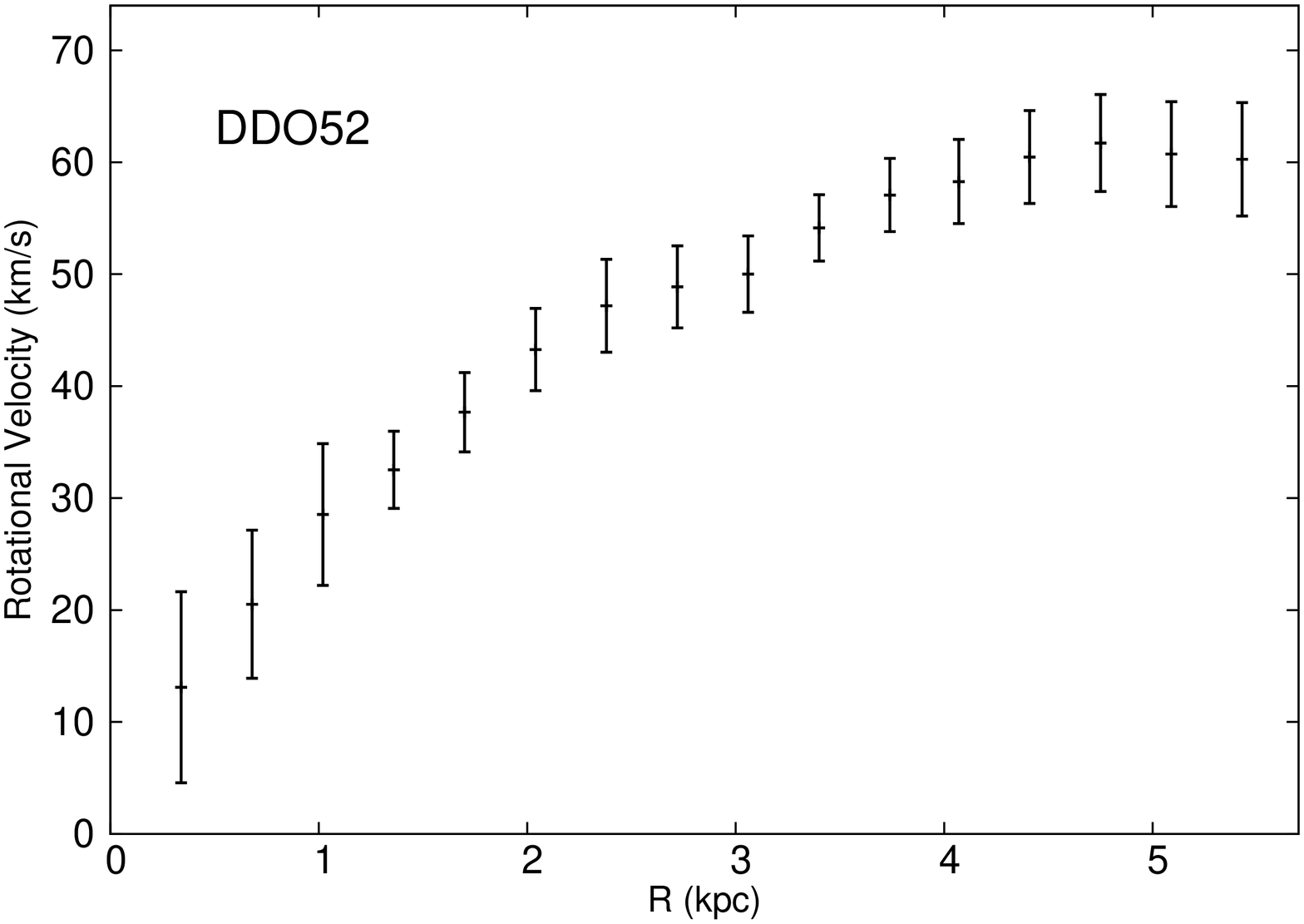,angle=270,width=7.05cm}}
\vskip -4.980cm
\hspace{-1.0cm}
\rightline{\epsfig{file=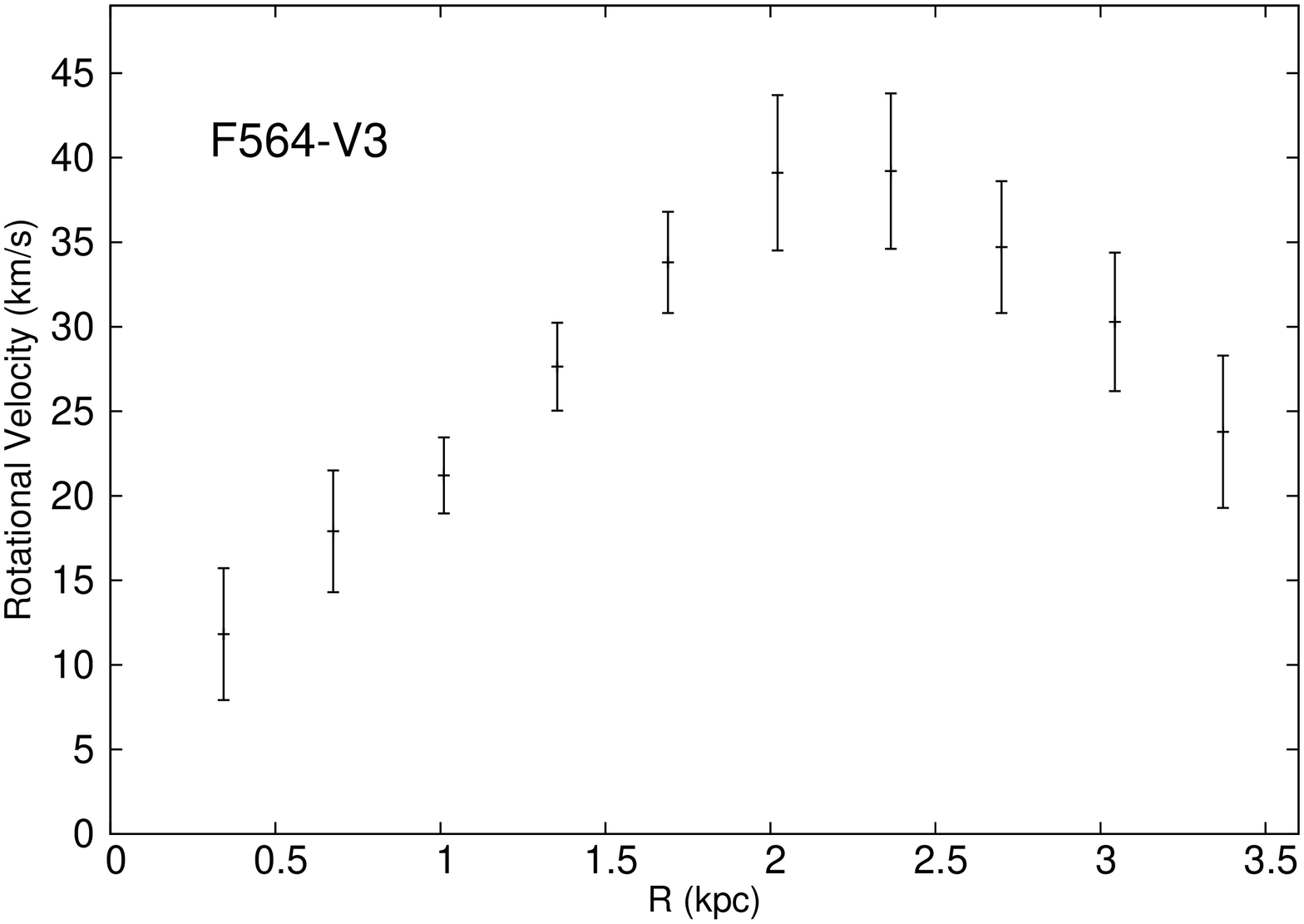,angle=270,width=7.05cm}}
\vskip 0.51cm
\hspace{0.11cm}
\leftline{\epsfig{file=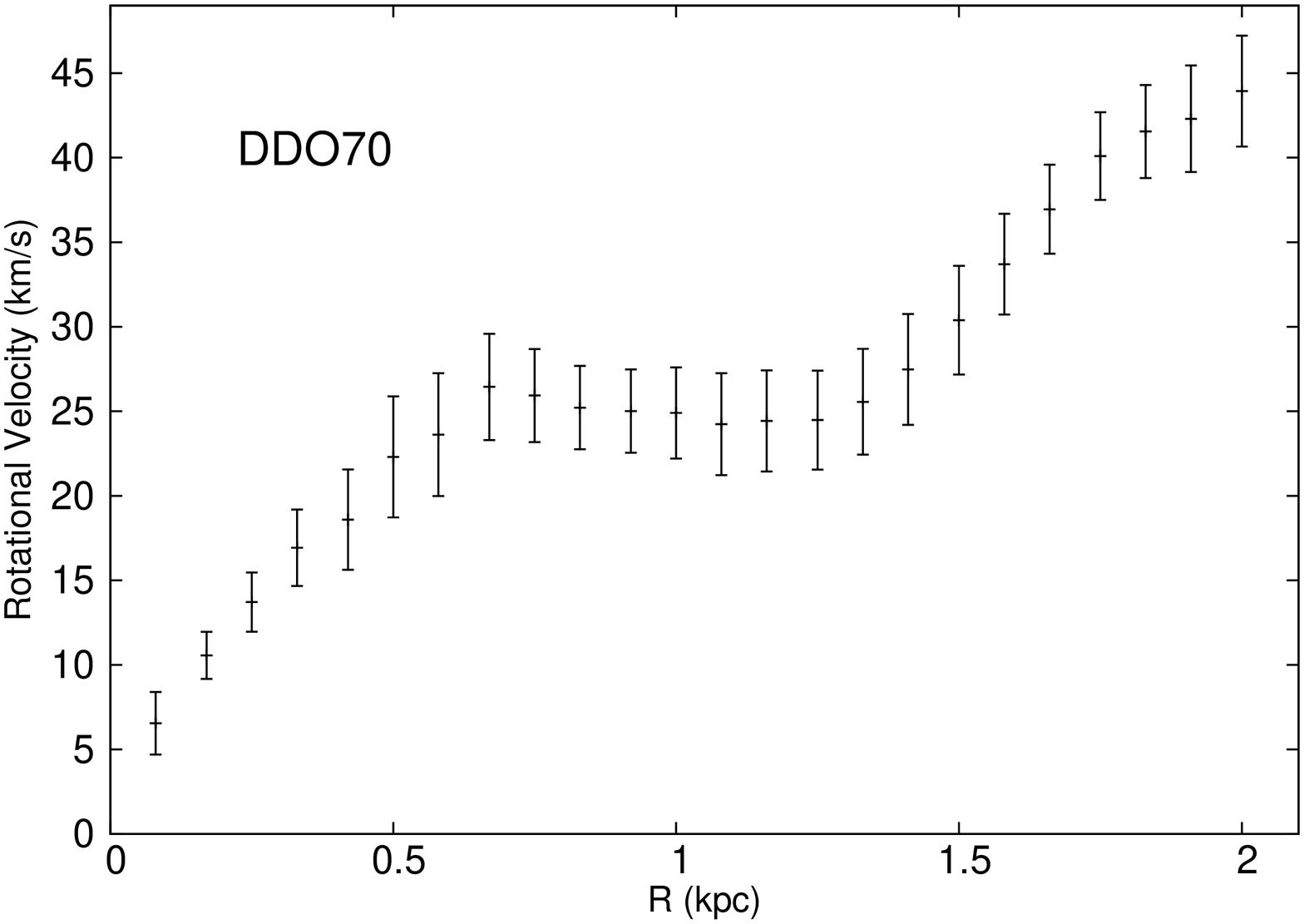,angle=270,width=7.05cm}}
\vskip -4.98cm
\hspace{-1.0cm}
\rightline{\epsfig{file=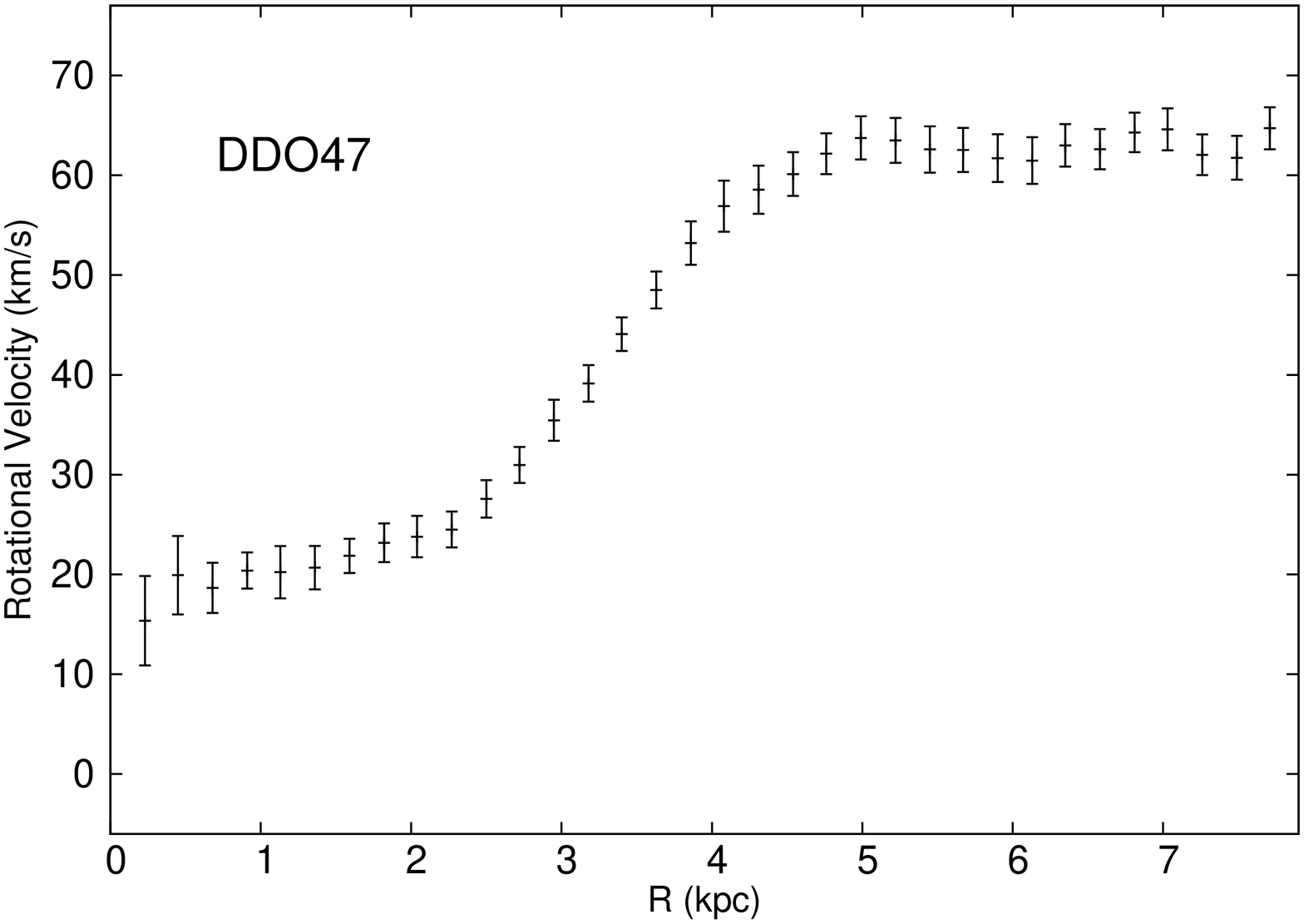,angle=270,width=7.05cm}}
\vskip 0.54cm
\noindent
{\small
Figure 1: Examples of the four classes of galaxies in the LITTLE THINGS sample: 
DDO52, classically shaped rotation curve (linear
rise, transiting to flat), F564-V3, classically shaped but with a downturn in $v_{rot}$,
DDO70, classically shaped but with a `hump' at large $R$,  
DDO47, an irregularly shaped rotation curve. 
}
\vskip 0.1cm

In the literature UV 
and $H\alpha$ emission have both been used as traces for star formation rates, see e.g. \cite{UVSFR,flux2,flux}.
The UV radiation is directly emitted from the photospheres of large stars, 
with $M_* \stackrel{>}{\sim} 3M_\odot$ (O- through later-type B-stars) 
while the $H\alpha$ nebular emission results from 
the recombination of hydrogen which can only be ionized by very large stars,
with $M_* \stackrel{>}{\sim} 17M_\odot$
(the most massive O- and early- type B-stars). 
The $H\alpha$ flux provides a measure of star formation rate on fairly short timescales of $\sim$ 10 Myr, while  
UV flux provides a measure on a somewhat larger timescales: $\sim$ 100 Myr. 
The timescale most relevant for the supernovae rate, $\Sigma_{SN}$,
in Eq.(\ref{3z}) is some appropriately weighted average over the timescale 
characteristic of the dissipative dynamics, $\tau \sim (3/2)n(r)T/\Gamma_{cool}$, expected to be less than
around a Gyr as mentioned earlier. 
A priori, both UV and $H\alpha$ fluxes present reasonable candidates for the relevant supernovae rate, and
we will consider both profiles in the $\chi^2$ analysis to follow. 

The Galaxy Evolution Explorer (GALEX) \cite{galex} has obtained images
of nearby galaxies in the far UV (FUV) bandpass of  1350-1750 $\AA$ and near UV (NUV) bandpass of 1750-2800 $\AA$.
The GALEX data was analyzed in  \cite{hunter} for a large selection of dwarf galaxies, including 
24 of the LITTLE THINGS sample of 26 galaxies, yielding the
azimuthally averaged surface brightnesses, $\mu_{FUV}(r)$. 
This quantity can be converted to the flux density via the standard formula: 
$F_{FUV}(r) = 10^{-0.4(\mu_{FUV}(r) + 48.6)} \ {\rm erg\ cm}^{-2}$ \ ${\rm s}^{-1}$ \ ${\rm Hz}^{-1}$ \ ${\rm arcsec}^{-2}$, 
and similarly for the NUV flux.
Also given in that reference is the $H\alpha$ surface brightnesses, $\mu_{H\alpha}$, 
which can also be converted to a $F_{H\alpha}(r)$ flux.
In the following, we shall model $\Sigma_{SN}(r)$ with each of these fluxes.
Of most interest
is the radial dependence of these quantities as the overall proportionality constant can be absorbed into
a redefinition of the coefficient $\lambda$ of Eq.(\ref{3z}); this redefinition is denoted as $\stackrel{\sim}{\lambda}$ (and will
be defined more precisely later on).

% xxxxx
Consider first the 18 classical dwarfs and DDO47.
In table 1 we give some properties of these galaxies: the distance of the galaxy as given in \cite{oh} and 
also the absolute AB FUV-band  magnitude (evaluated from the apparent magnitude assuming this distance measurement). \footnote{For dwarf 
galaxies extinction corrections are generally small. 
For the estimated FUV absolute magnitude of the dwarf galaxies the internal and galactic (foreground) extinction corrections were 
nevertheless included following the procedure 
of  \cite{flux,flux2} using data obtained by the online updated nearby galaxy catalogue \cite{table}.}
Also given in table 1 is  $\chi^2_r = \chi^2/N_{dof}$ 
(for a 1 parameter fit, the number of degrees of freedom is $N_{dof} = N_{data}-1$) 
for the fit of the model to the rotation curve data, modelling $\Sigma_{SN}(r)$ with a) $F_{FUV}(r)$, b) $F_{NUV}(r)$ and
c) $F_{H\alpha}$, each obtained from \cite{hunter} as described above.
\footnote{For the dwarfs: DDO52, DDO101, DDO210, DDO216. 
 there were insufficient  H$\alpha$ flux measurements for which to construct the flux, while for IC10 UV
flux measurements were unavailable.}
%%% moved here xxxx%%%
Here, the $\chi^2$ function for each galaxy is constructed in the usual way:
\begin{eqnarray}
\chi^2 \equiv \sum_{i=1}^{N_{data}} \left[ {v_{rot}(i) - v_{rot}^{exp}(i) \over \delta v_{rot}^{exp}(i)} \right]^2
\end{eqnarray}
where the $N_{data}$ (asymmetric drift corrected) binned rotation curve measurements, 
$v_{rot}^{exp}(i)\pm \delta v_{rot}^{exp}(i)$, are obtained from \cite{oh} and
$v_{rot}$ is evaluated via Eq.(\ref{rotx}) (with $v_{gas}$ and $v_{stars}$ from \cite{oh}).
This $\chi^2$ function was minimized with respect to variations of the parameter
$\lambda$ (which was allowed to have a different value for each galaxy).
The related quantity, $\stackrel{\sim}{\lambda}$, to be defined in Eq.(\ref{10x}), is also given in the table.

The $\chi^2$ values given in the table suggests that the density profile, Eq.(\ref{3z}), motivated by the dissipative dynamics
is consistent with 
a zeroth order approximation for most of the 18 classical dwarfs.   
The resulting fits are shown in figures 2-4 for the 18 `classical'
dwarfs and DDO47. 
In these figures, the FUV flux was used as the tracer for $\Sigma_{SN}$; very similar results 
occur for the other two traces considered. Indeed, as the $\chi_r^2$ values in the table indicate, use of the NUV flux generally leads to a
slightly better fit to the rotation curve data, while with the $H\alpha$ flux the fit fares slightly worse. 
It is also noted that use of UV flux generally leads to a slightly better fit to the rotation curves than
the Kennicutt-Schmidt-type power law considered in \cite{foot5}.\footnote{A curious exception is DDO47, for which the Kennicutt-Schmidt type
power law ($\Sigma_{SN}(r) \propto [\Sigma_{gas} (r)]^N$) leads to a significantly better fit: 
$\chi^2_r (KS) = 5.2$ (for exponent $N=2$), $\chi^2_r (KS) = 6.7$ (for exponent $N=3$) 
cf. $\chi^2_r$(FUV) = 9.7.}

The density profile, Eq.(\ref{3z}), is highly constrained as it depends on just one parameter $\lambda$.
This
parameter sets the normalization of the dark halo contribution of the rotation curve.
The shape of the curve is fully predicted by Eq.(\ref{3z}) once the supernovae distribution is input via
the FUV flux radial profile. Figures 2-4 indicate that the radius where the rotation curve transitions from 
linear to flat is generally reproduced reasonably
well. 
\footnote{
In fact, this reproduction is also insensitive to the uncertainty in the distance of the galaxy, $D$. 
This can most easily be seen by working with the angular variable, $r/D$, and noting that both rotation curves and surface photometry
are measured in terms of such a  quantity.
}
Note though that the rotation curve error bars are not purely statistical as they are usually derived by
taking the difference between rotation curve measurements from each side of the galaxy. It is possible that the small but 
perceptible difference
between the fit and the data of some of these galaxies
is an indication for corrections
to the zeroth order approximation of Eq.(\ref{3z}).
There are many possible sources of corrections including: radial dependence of the halo temperature, modelling of the supernova distribution, 
correcting for azimuthal asymmetry etc.  
A simple way to incorporate first order corrections is 
%assume the dark matter density of the form Eq.(\ref{3z}) 
to allow  $\lambda$ in Eq.(\ref{3z}) to be spatially dependent: $\lambda (r) = \lambda_0 + \lambda_1 r + ...$
Keeping only the first two terms gives a dark matter density 
defined in terms of two parameters, $\lambda_0, \ \lambda_1$.
The two parameter fit for two examples, DDO154 and WLM
is shown in figure 5. As this figure illustrates, the inclusion of first order corrections in this simple
phenomenological manner allows for an improved representation of the data.

\begin{table}
\centering
\begin{tabular}{c c c c c c c }
\hline\hline
Galaxy & D (Mpc) &  $M_{FUV}$ & $\chi^2_r$ (FUV) & $\chi^2_r$ (NUV) & $\chi^2_r$ ($H\alpha$) & $\stackrel{\sim}{\lambda}$ 
(km$^2$/s$^2$) 
{\rule{0pt}{2.9ex}}       
{\rule[-1.5ex]{0pt}{0pt}} 
\\ [0.5ex]
\hline
  & \multicolumn{4}{c}{\ \ \ \ \ \ \ \ \ \ \ \ 18 classical dwarfs }  
{\rule{0pt}{2.9ex}}       
{\rule[-1.5ex]{0pt}{0pt}} 
\\ 
\hline
DDO43 & 7.8 & -13.14 & 0.55 & 0.69 & 0.72 & \ 5.32E-3 
{\rule{0pt}{2.9ex}}       
\\
DDO50 & 3.4 & -15.42 & 3.53 & 3.52 & 3.06 & 4.36E-4
\\
DDO52 & 10.3 & -13.36 & 0.18 & 0.15 & - & 1.50E-2
\\
DDO53 & 3.6 & -12.51 & 1.30 & 1.25 & 1.30 & 4.26E-3
\\
DDO87 & 7.7 & -13.09 & 0.96 & 0.79 & 2.54 & 1.41E-2
\\
DDO101 & 6.4 & -11.59 & 28.6 & 22.9 & - & 0.11   
\\
DDO126 & 4.9 & -13.39 & 0.56 & 0.55 & 0.92 & 4.62E-3 
\\
DDO133 & 3.5 & -13.00 & 5.1 & 4.1 & 6.4 & 1.22E-2
\\
DDO154 & 3.7 & -13.10 & 0.81 & 0.83 & 0.80 & 1.01E-2
\\
DDO210 & 0.9 & -8.23 & 0.84 & 0.69 & - & 4.53E-2
\\
DDO216 & 1.1 & -9.48 & 0.50 & 0.53 & - & 2.98E-2
\\
IC10 & 0.7 & - & - & - & 0.07 & -
\\
NGC1569  & 3.4 & -16.80 & 0.61 & 0.61 & 0.50 & 9.81E-5 
\\
NGC2366  & 3.4 & -15.32 & 0.28 & 0.27 & 0.23 & 1.90E-3
\\
%NGC3738  & 4.9 & -14.72 & 6.14 & 2.96 for $r < 1.2$
NGC3738  & 4.9 & -14.72 & 6.14 & 4.27 &  3.32 & 1.56E-2
\\
WLM  & 1.0 & -12.51 & 0.53 & 0.44 & 0.92 & 9.29E-3
\\
Haro29  & 5.9 & -13.68 & 0.40 & 0.40 & 0.55 & 3.71E-3
\\
Haro36  & 9.3 & -14.61 & 1.86 & 1.63 & 1.90    & 1.72E-3
\\ [0.3ex]
\hline
 & \multicolumn{4}{c}{
\ \ \ \ \ \ A dwarf with irregularly shaped rotation curve 
} 
{\rule{0pt}{2.9ex}}       
{\rule[-1.5ex]{0pt}{0pt}} 
\\ 
\hline
DDO47 & 5.2 & -13.83 & 9.7 & 10.8 & 12.5 & 9.33E-3 
{\rule{0pt}{2.9ex}}       
{\rule[-1.5ex]{0pt}{0pt}} 
\\
\hline\hline
\end{tabular}
\vskip 0.3cm
\caption{
{\small
LITTE THINGS classical dwarfs (defined in text) and DDO47.
$D$ ($M_{FUV}$) is the distance (FUV absolute AB magnitude) of the dwarf
and  
$\chi^2_r$ given for the 1-parameter ($\lambda$) fit of the rotation curve data to the model,
Eq.(\ref{4z}), Eq.(\ref{3z}),
modelling $\Sigma_{SN}$ with a) FUV, b) NUV and c) $H\alpha$ surface brightness radial profiles.
The fitted parameter $\stackrel{\sim}{\lambda}$, Eq.(\ref{10x}), is also given.
}
}
\end{table}

\begin{table}
\centering
\begin{tabular}{c c c  c c }
\hline\hline
Galaxy & D (Mpc) &  $M_{FUV}$ &  $\chi^2_r$ with $R^*$ (kpc) & $\stackrel{\sim}{\lambda}$  
(km$^2$/s$^2$) 
{\rule{0pt}{2.9ex}}       
\\ [0.4ex]
\hline
% & \multicolumn{3}{c}{\ \ \ \ \ Three dwarfs with downturn in $v_{rot}$}
%\T
%\B
%\\ [0.3ex]
%\hline
DDO46 & 6.1 & -13.15 &   3.85 with $R^* = 1.90$ & \ 3.19E-2
{\rule{0pt}{2.9ex}}       
\\
DDO168 & 4.3 & -13.64 &  3.20 with $R^* = 3.07$ & 9.10E-3
\\
F564-V3 & 8.7 & -10.94 &  0.56 with $R^* = 2.15$ & \ 5.47E-2
{\rule[-1.5ex]{0pt}{0pt}} 
\\
\hline\hline
\end{tabular}
\vskip 0.3cm
\caption{
{\small
LITTLE THINGS dwarfs with a downturn in the rotational velocity. 
$D$ ($M_{FUV}$) 
is the distance 
(FUV absolute AB magnitude)  
of the dwarf and 
$\chi^2_r$ is given for the 2-parameter ($\lambda$, $R^*$) fit of the rotation curve data to the model,
Eq.(\ref{4z}), Eq.(\ref{3z}), Eq.(\ref{7xy}).
The supernovae distribution, $\Sigma_{SN}$, is here modelled with the FUV surface brightness radial profile.
The fitted parameter $\stackrel{\sim}{\lambda}$, Eq.(\ref{10x}), is also given.
}
}
\end{table} 
% put table 1 here 

We now consider the three dwarfs with a downturn in $v_{rot}$.
A possible explanation for the downturn is that it represents the physical extent of the dark halo for these galaxies. 
If this is the case then we could try to model this boundary effect with
a sharp cut-off:
\begin{eqnarray}
\lambda (r) &=& \lambda \ \ {\rm for}\  r \le R^*, 
\nonumber \\
\lambda (r) &=& 0 \ \ {\rm for}\  r >  R^* \ .
\label{7xy}
\end{eqnarray}
This introduces one additional parameter, $R^*$.
The $\chi^2_r$ values for this two parameter ($\lambda, \ R^*$) fit are given in table 2 and the resulting fits are shown
in figure 6. 
It is of course also  possible that the downturn
is an indication that the halos of these dwarfs are not currently in a steady-state configuration
due to some perturbation. This is also a possible explanation of the dwarfs with a `hump', as we will now
examine.

The four remaining dwarfs of the LITTLE THINGS sample are those which feature a `hump' at the 
boundary region. The hump feature 
might be taken as an indication that the halos of these dwarfs are not entirely in a steady-state
configuration where heating and cooling rates locally balance. This situation 
could conceivably arise  if a galaxy was 
being perturbed, either externally by a nearby
galaxy or via some internal mechanism.  Such perturbations, if significant, would affect the star
formation rate of the galaxy. This suggests that a galaxy's recent star formation history could 
possibly be used as a diagnostic aid in determining whether its dark halo is likely to 
be in a steady-state configuration.

It may also be possible for a galaxy to be in a partially equilibrated state. Perhaps  
only the inner part of the halo
could be in an approximate steady-state configuration.  
This situation might arise because of the disparate timescales:
the higher density inner region of a dark halo can cool and equilibrate more rapidly than 
the less dense outer region.  
The dwarfs with a `hump' (and possibly also those with a `bump') might result from such a condition.
In table 3 and figure 7 we show the results of the fit to the inner region of the dwarfs with a `hump'.

It is perhaps noteworthy that some of the dwarfs which feature poor fits 
to the steady-state solution, Eq.(\ref{3z}), are known to be undergoing starburst activity. 
In particular, of the four dwarfs with the lowest $\lambda$ values, DDO50, NGC1569, IC1613 and CVn1dwA, three of them
[NGC1569, DDO50 (Holmberg II) and CVn1dwA (UGCA292)] are known to be in a starburst phase, e.g. \cite{flux,starburst}.
In fact, a low value of $\lambda$ is an indication that
halo heating exceeds cooling ($\Gamma_{heat} > \Gamma_{cool}$) a situation which would imply that these halos are currently 
expanding.

Dissipative dynamical halos of the type envisaged here would be 
expected to play an important role in regulating the star formation rate of a given galaxy.
Any perturbation either external or internal that causes
the halo to move out of equilibrium, $\Gamma_{heat} \neq \Gamma_{cool}$,
would lead to an expanding or contracting halo dynamically adjusting under the various forces.
Such dynamics, which could in principle be described by Euler's equations of fluid dynamics, 
would be expected to be a major influence on ordinary star formation rates.
As the halo expands or contracts, the baryonic
gas density decreases or increases in response to the changing
gravity. 
These physical changes in baryonic gas density would in turn correlate with the star formation rate (as suggested by
the Kennicutt-Schmidt relation). 
This interplay between the dynamical halo and star formation rates could be one of the main drivers of starburst activity.
It is a temporary non-equilibrium phase where $\Gamma_{cool} \neq \Gamma_{heat}$ and the halo is undergoing significant 
changes influencing the star formation rate and indeed vice versa.

\begin{table}
\centering
\begin{tabular}{c c c  c c }
\hline\hline
Galaxy & D (Mpc) &  $M_{FUV}$ & $\chi^2_r$ (FUV) for $r < R_1$ (kpc)  & $\stackrel{\sim}{\lambda}$ 
(km$^2$/s$^2$) 
{\rule{0pt}{2.9ex}}       
{\rule[-1.5ex]{0pt}{0pt}} 
\\ 
\hline
% & \multicolumn{3}{c}{Five dwarfs with a `hump' in $v_{rot}$}
%\T \B
%\\ [0.4ex] \hline
CVn1dwA & 3.6 & -11.56 & 1.24 for $r < 1.4$ & \ 1.41E-3
{\rule{0pt}{2.9ex}}       
\\
DDO70 & 1.3 & -11.92 & 0.71 for $r < 1.6$ & 1.19E-2
\\
IC1613 & 0.7 & -12.80 & 1.15 for $0.3 < r < 1.7$ & 2.38E-4
\\
UGC8508  & 2.6 & - & 1.72 for $r < 1.4$ & -
{\rule[-1.5ex]{0pt}{0pt}} 
\\
\hline\hline
\end{tabular}
\vskip 0.2cm
\caption{
{\small
LITTLE THINGS dwarfs with a `hump'  in the rotational velocity. 
$D$ ($M_{FUV}$)
is the distance (FUV absolute AB magnitude) and 
$\chi^2_r$ is given for the 1-parameter ($\lambda$) 
fit of the rotation curve data to the model,
Eq.(\ref{4z}), Eq.(\ref{3z}), for the inner radial region, $r < R_1$.
The supernovae distribution, $\Sigma_{SN}$, is here modelled with the FUV surface brightness radial profile, 
except for UGC8508 which was modelled with $H\alpha$ (FUV was unavailable in the case).
The fitted parameter $\stackrel{\sim}{\lambda}$, Eq.(\ref{10x}), is also given.
}}
\end{table}

%%% paragraph about DDO101 %%%%%%%%%

The dwarf DDO101 has the distinction of featuring the worst fit, with $\chi^2_r (FUV) = 28.6$.
It was emphasised in \cite{oh} that the inner slope of DDO101 does have 
potentially large uncertainty due to possible beam smearing 
effects.
Whether such effects could reconcile the predicted rotation curve with the data is unclear.
If this (or other) systematic effects are not responsible for the disagreement, 
then one could speculate that this very faint dwarf has
evolved to the point where there is insufficient heating to support the halo.  The halo 
may have cooled and collapsed (or is in the process of collapsing) into dark stars. 
Indeed table 1 indicates that the
$\lambda$ value for this dwarf is anomalously large, which 
is evidence in favour of this interpretation.
%[A relatively large value of $\lambda$ is an indication that
%halo cooling exceeds heating, $\Gamma_{cool} > \Gamma_{heat}$.]

\newpage

%\vskip 0.3cm
\hspace{0.11cm}
%\centerline{\epsfig{file=fig1.eps,angle=0,width=11.0cm}}
\leftline{\epsfig{file=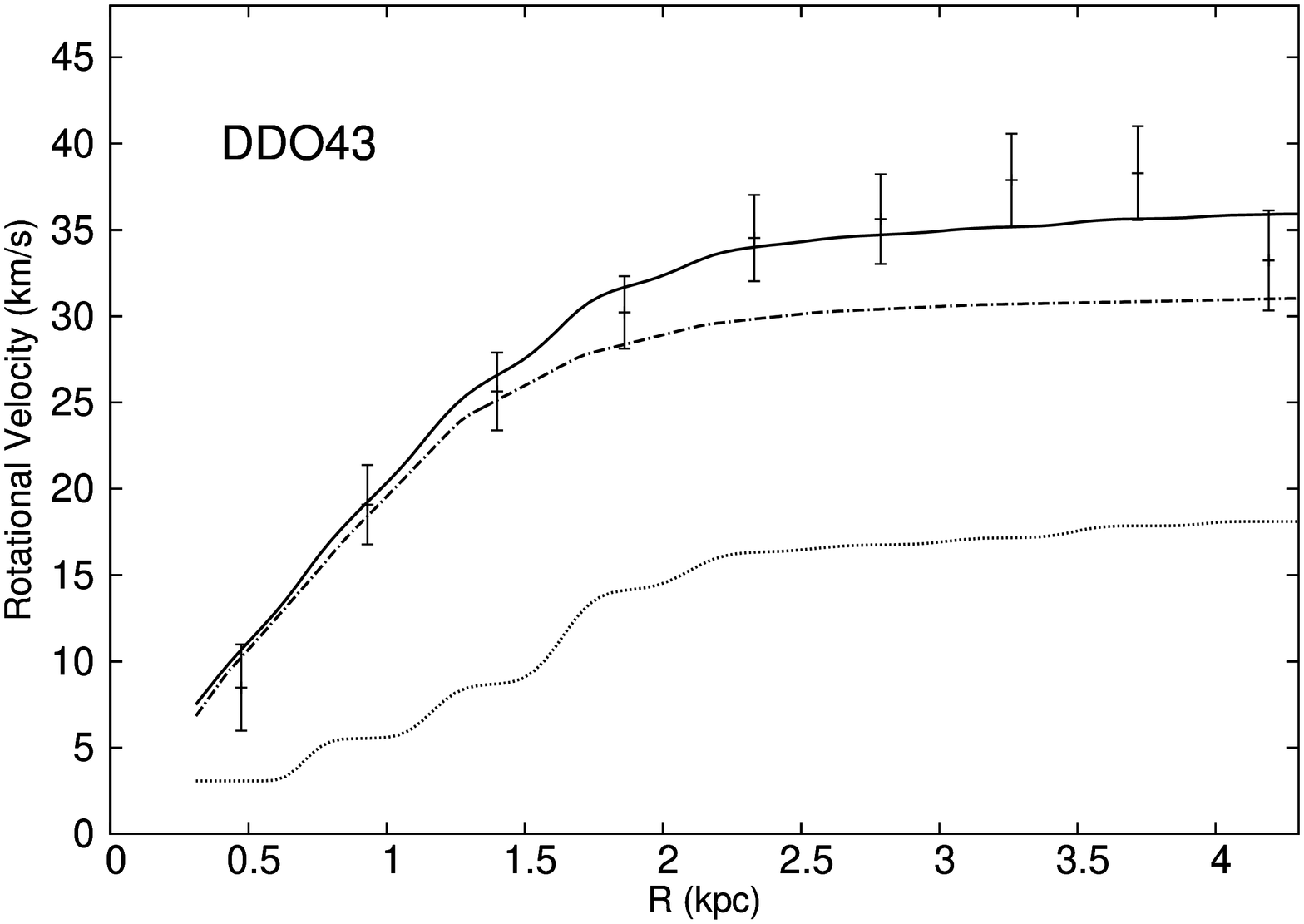,angle=270,width=7.05cm}}
\vskip -4.98cm
\hspace{-1.1cm}
\rightline{\epsfig{file=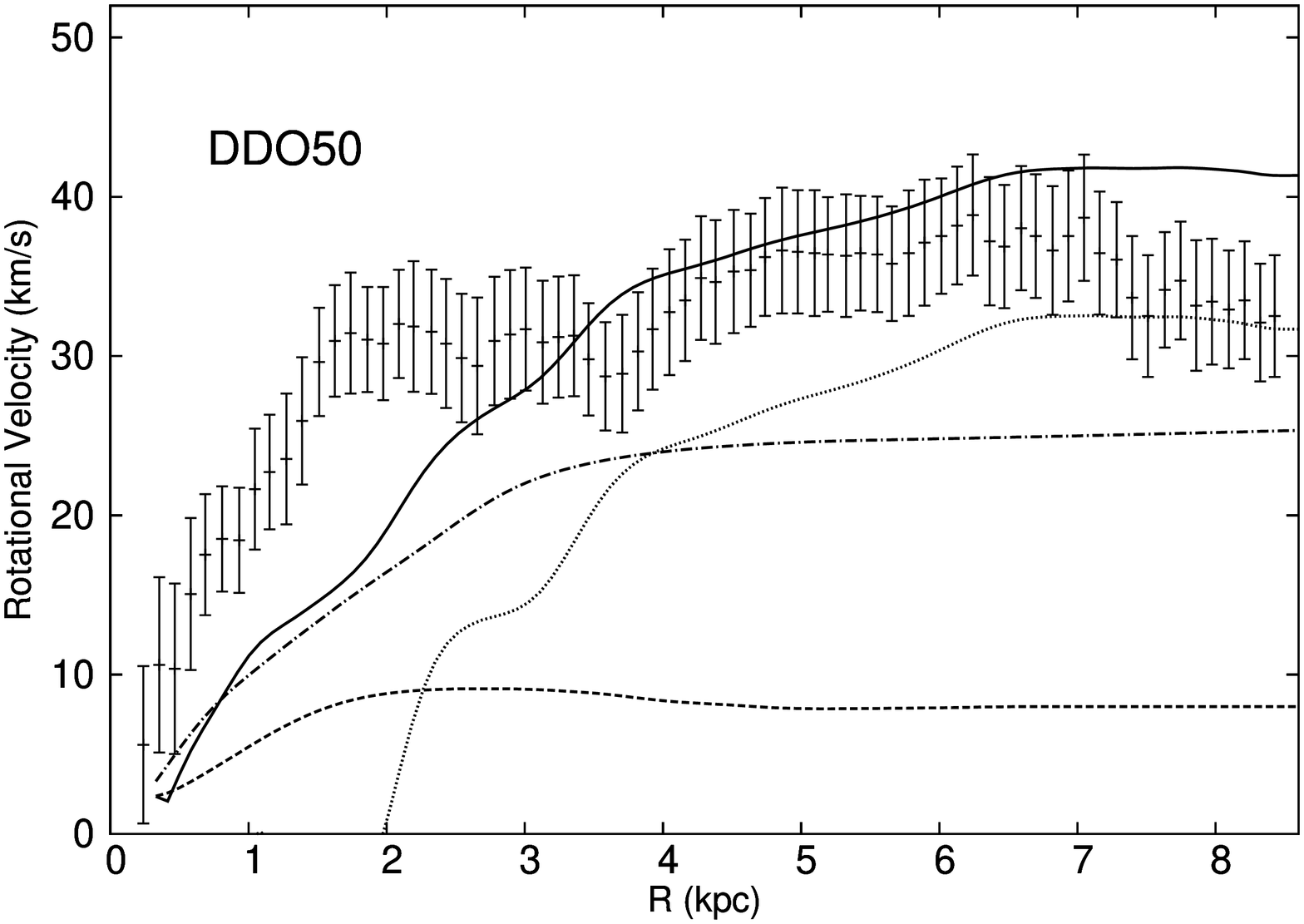,angle=270,width=7.05cm}}
\vskip 0.46cm
\hspace{0.11cm}
\leftline{\epsfig{file=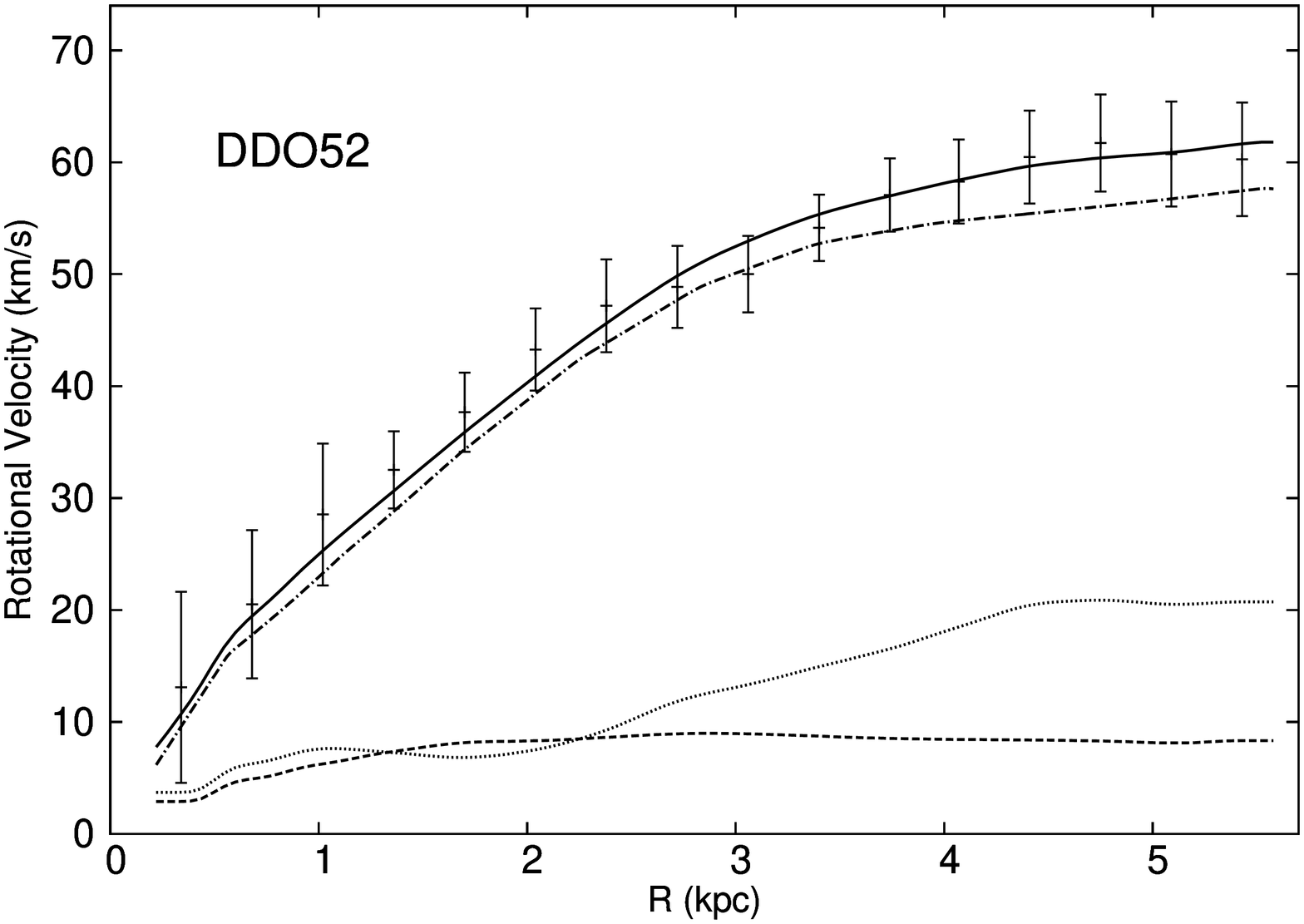,angle=270,width=7.05cm}}
\vskip -4.98cm
\hspace{-1.1cm}
\rightline{\epsfig{file=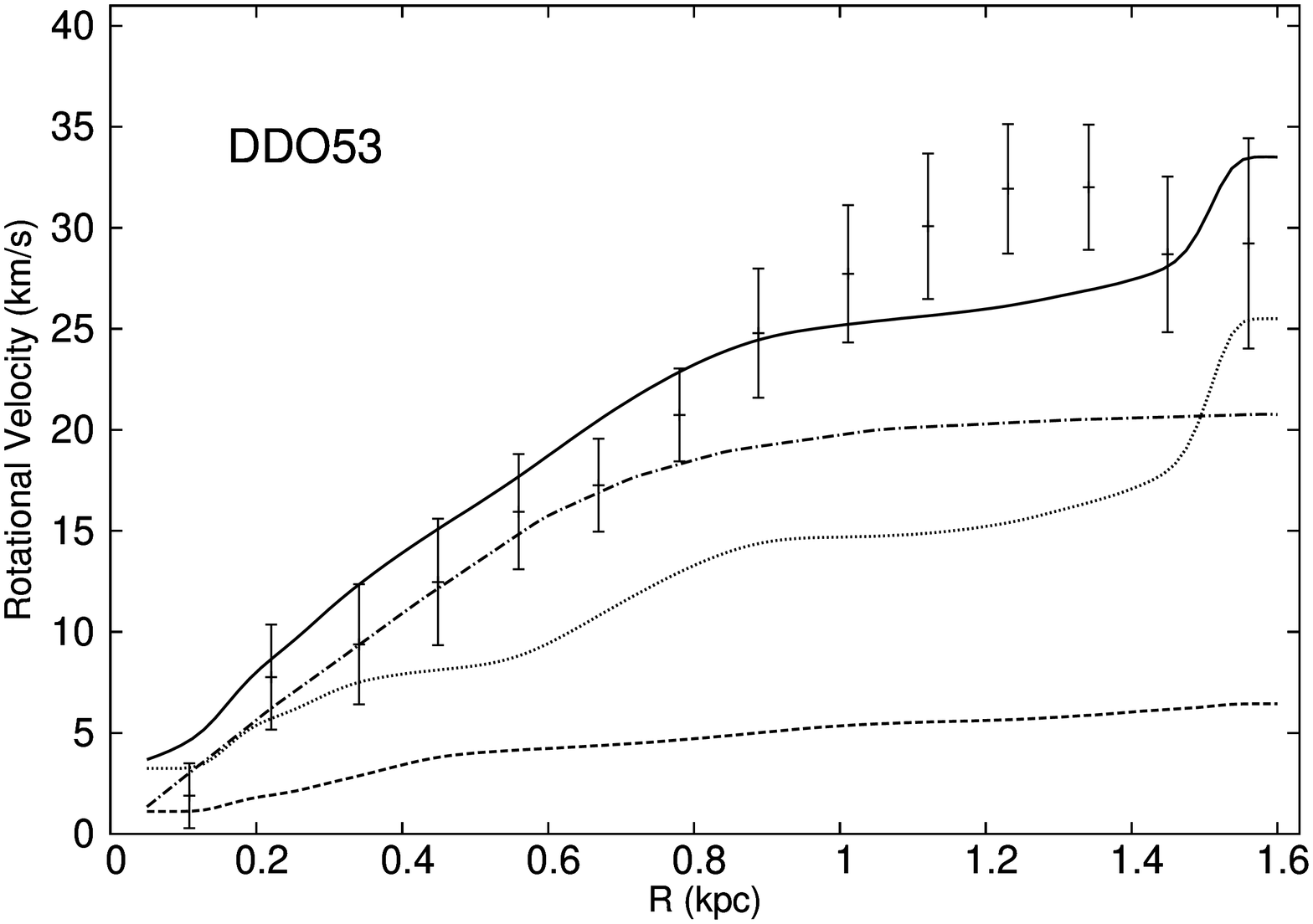,angle=270,width=7.05cm}}
\vskip 0.46cm
\hspace{0.11cm}
\leftline{\epsfig{file=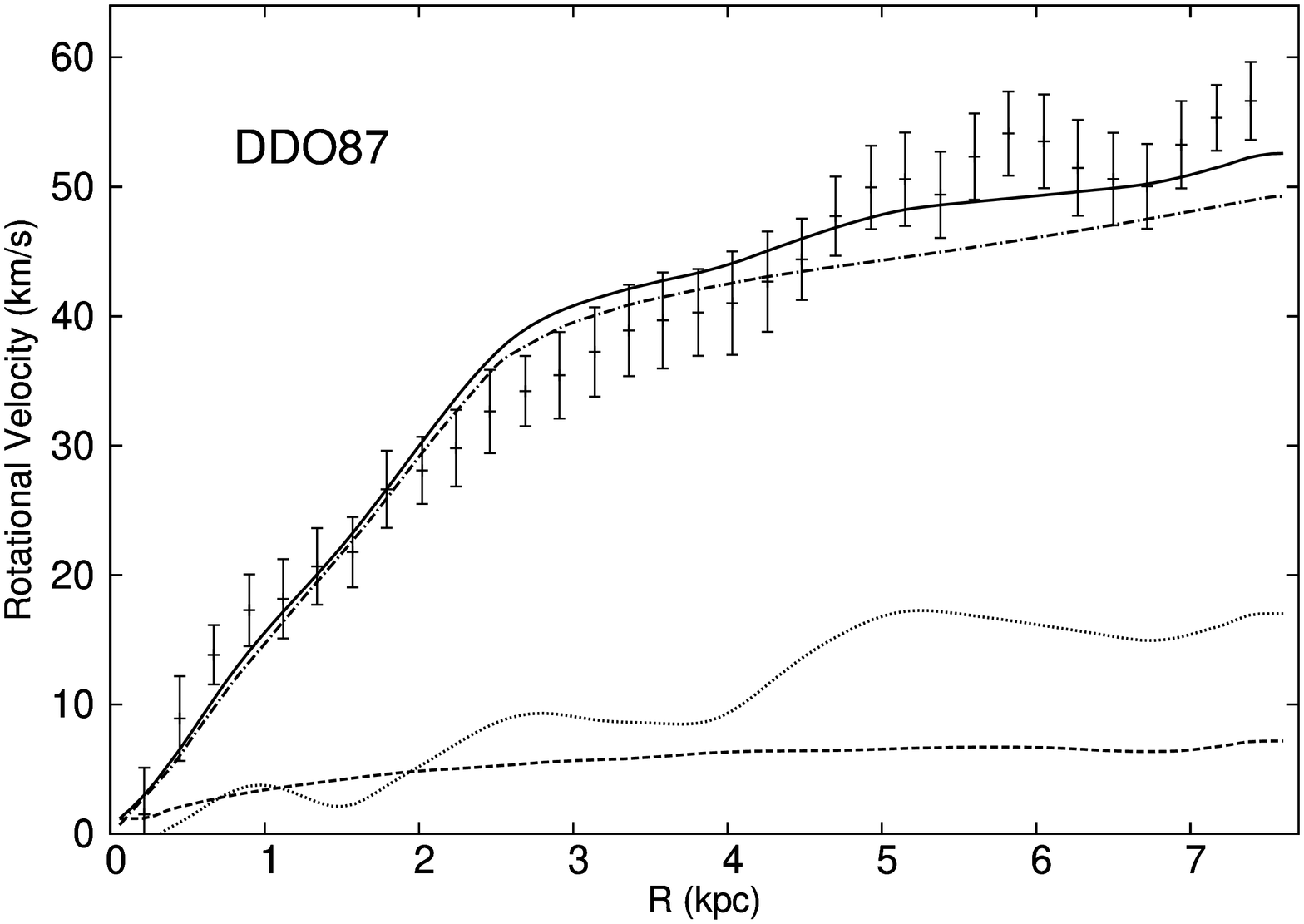,angle=270,width=7.05cm}}
\vskip -4.98cm
\hspace{-1.1cm}
\rightline{\epsfig{file=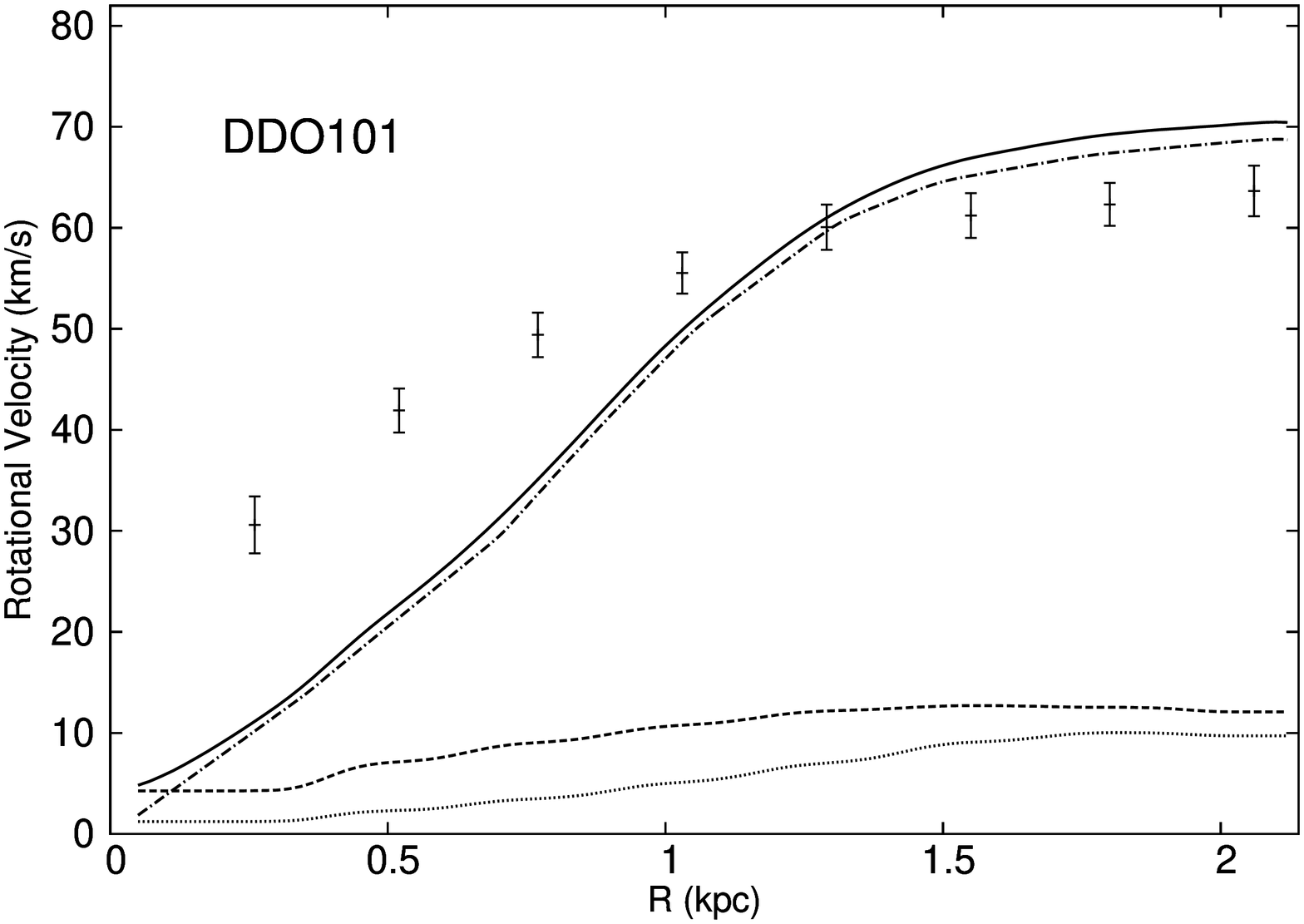,angle=270,width=7.05cm}}
\vskip 0.46cm
\hspace{0.11cm}
\leftline{\epsfig{file=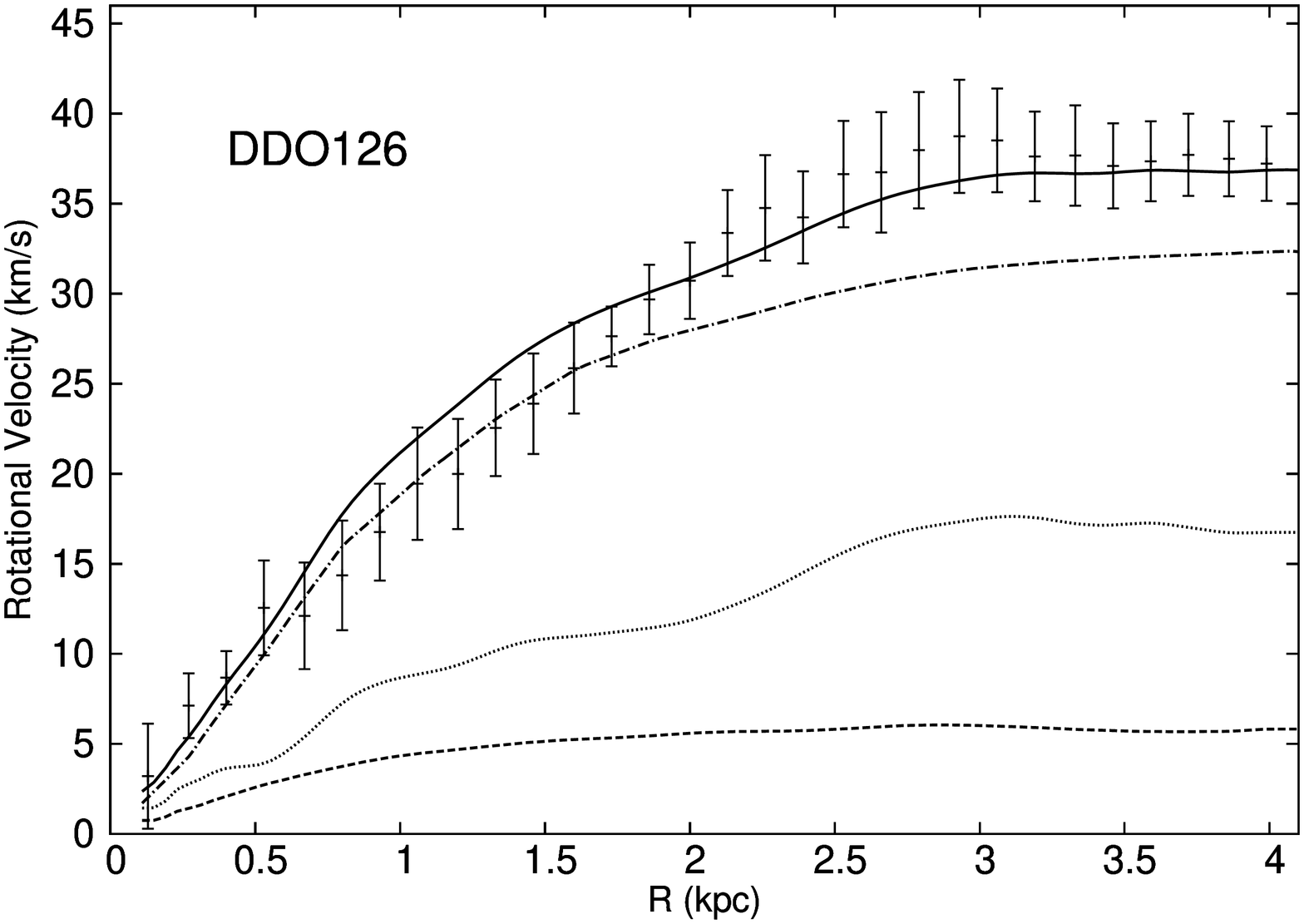,angle=270,width=7.05cm}}
\vskip -4.98cm
\hspace{-1.1cm}
\rightline{\epsfig{file=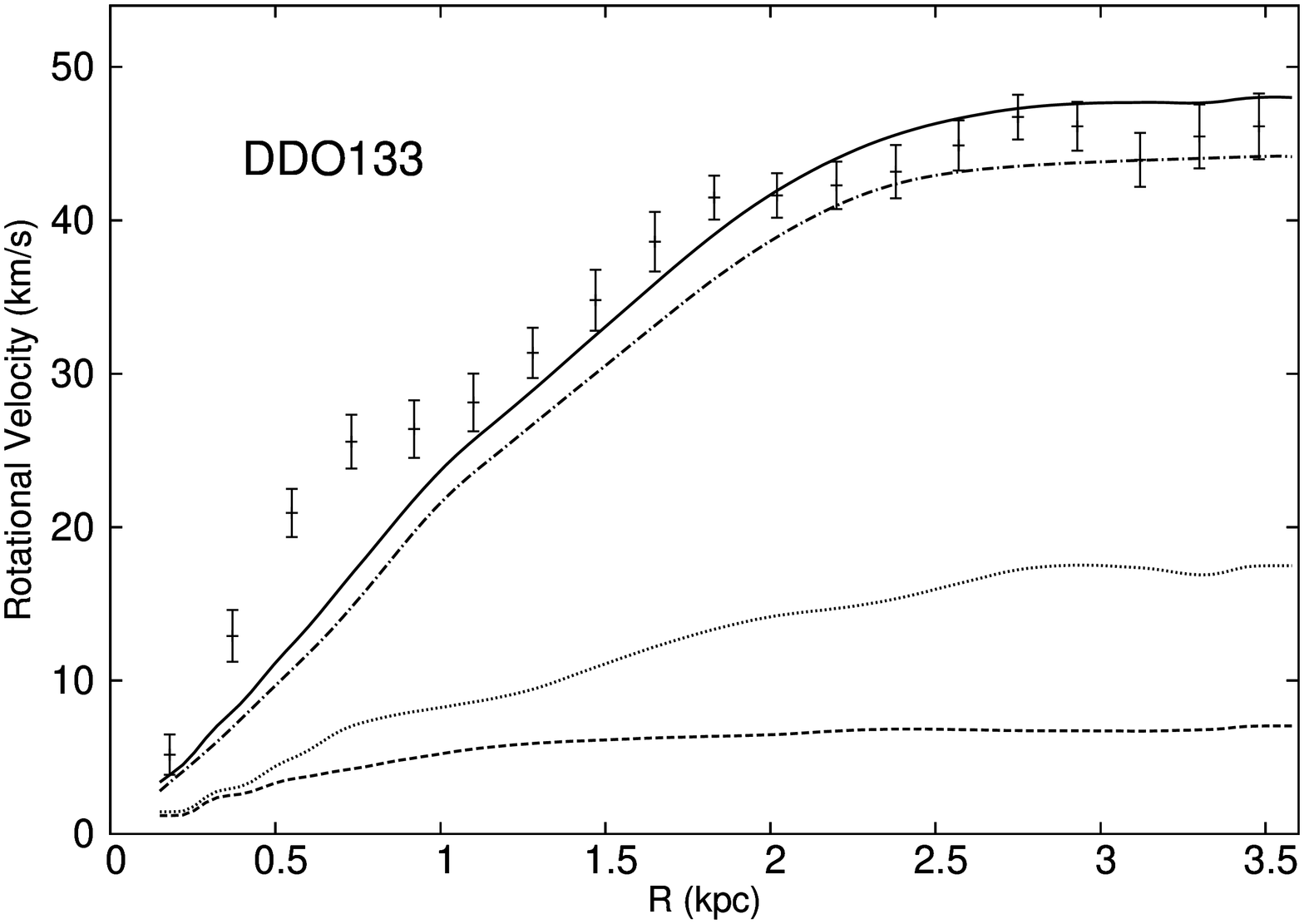,angle=270,width=7.05cm}}
\vskip 0.50cm
\noindent
{\small Figure 2: Rotation curves for the dwarf galaxies, DDO43, DDO50, DDO52, DDO53, DDO87,
DDO101, DDO126 and DDO133. The data is from \cite{oh}.
The stellar (dashed line), baryonic gas (dotted line) and dissipative dark halo (dashed dotted line) contributions
are shown. The solid line is the sum of these contributions added in quadrature.
The halo contribution results from the 1-parameter fit, Eq.(\ref{3z}) with $\lambda$ assumed spatially independent.
[$\Sigma_{SN}$ in Eq.(\ref{3z}) is modelled via the FUV surface brightness radial profile from \cite{hunter} as described in the text.]
}

\vskip 0.8cm
\hspace{0.11cm}
%\centerline{\epsfig{file=fig1.eps,angle=0,width=11.0cm}}
\leftline{\epsfig{file=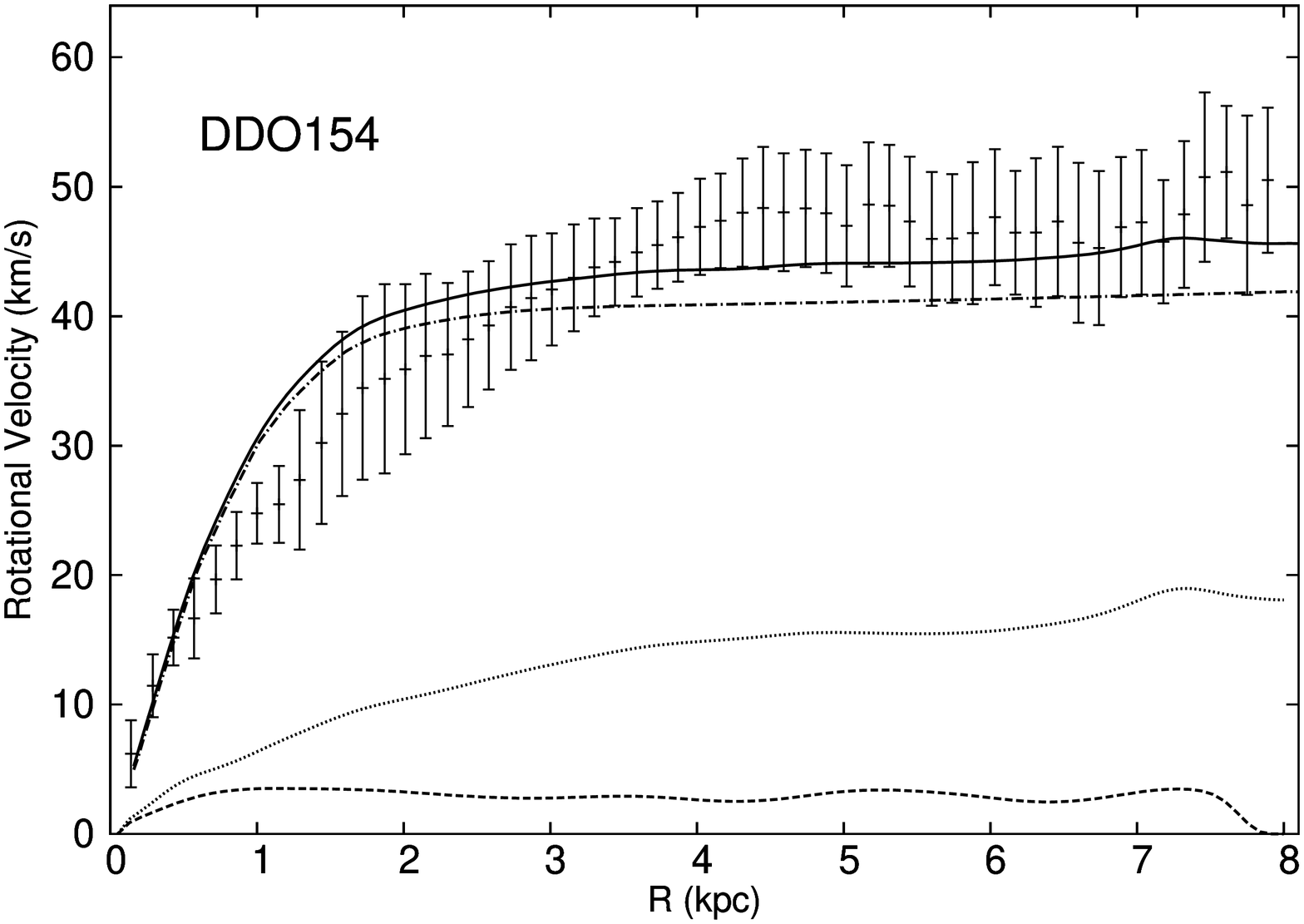,angle=270,width=7.05cm}}
\vskip -4.98cm
\hspace{-1.1cm}
\rightline{\epsfig{file=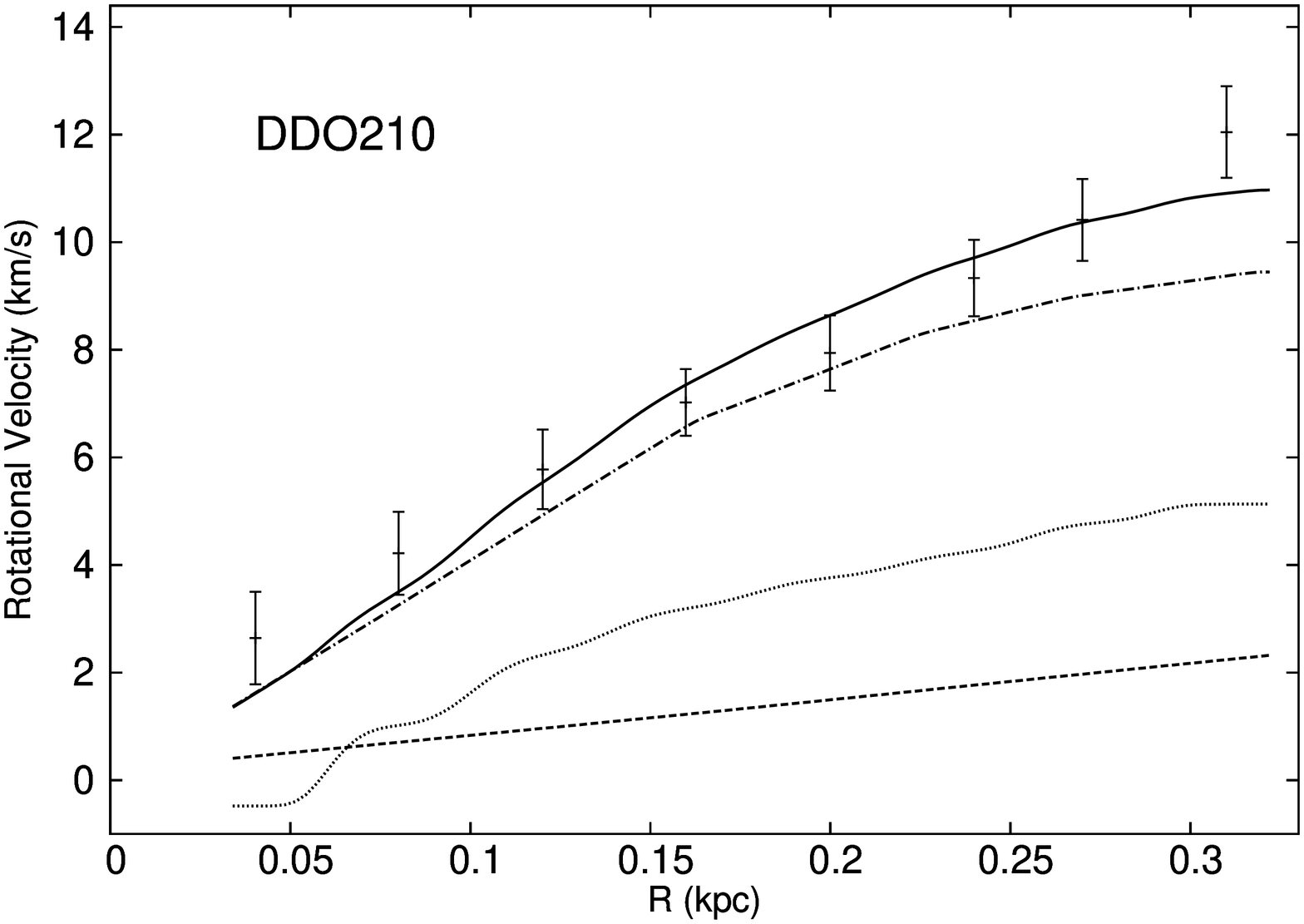,angle=270,width=7.05cm}}
\vskip 0.53cm
\hspace{0.11cm}
\leftline{\epsfig{file=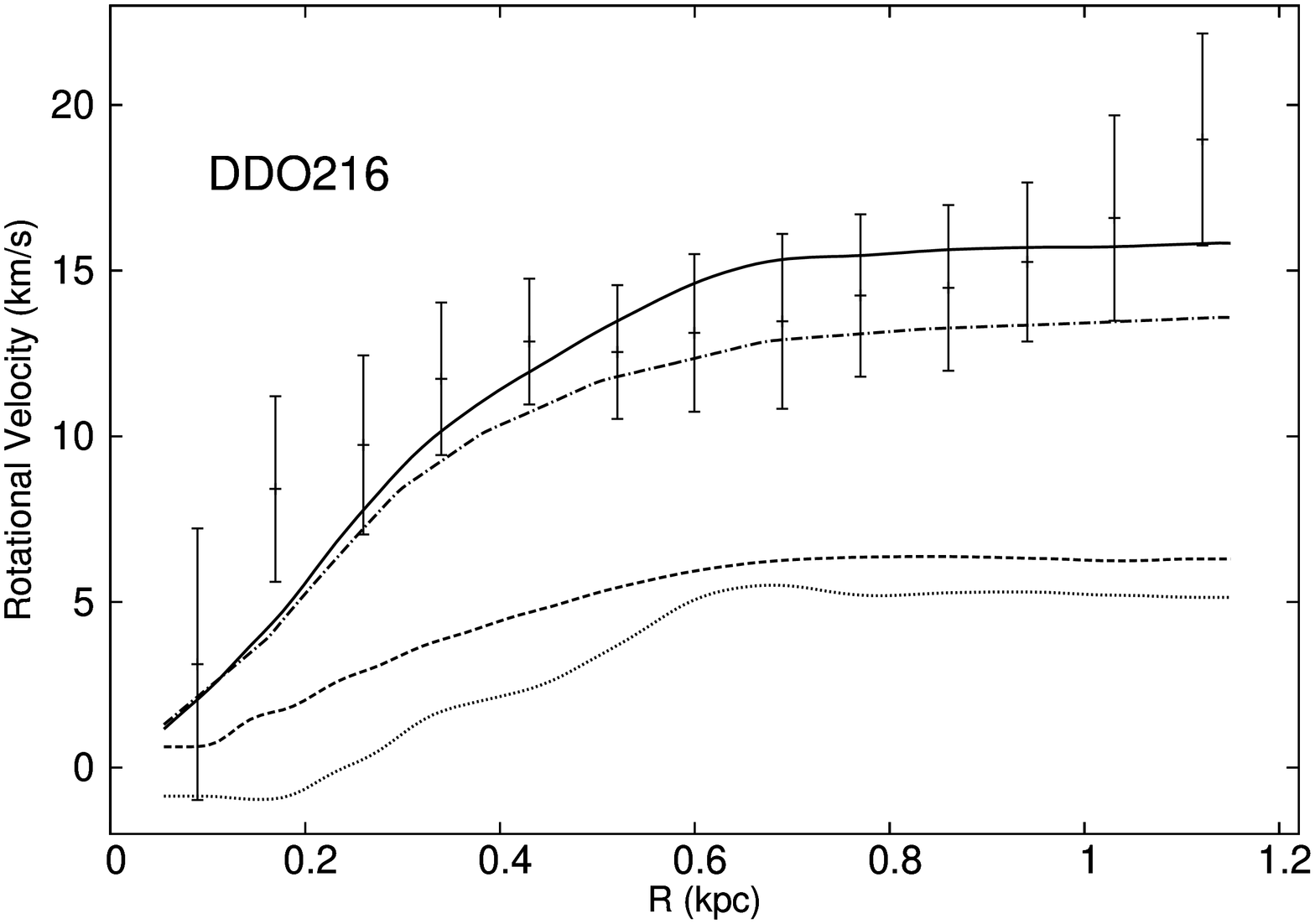,angle=270,width=7.05cm}}
\vskip -4.98cm
\hspace{-1.1cm}
\rightline{\epsfig{file=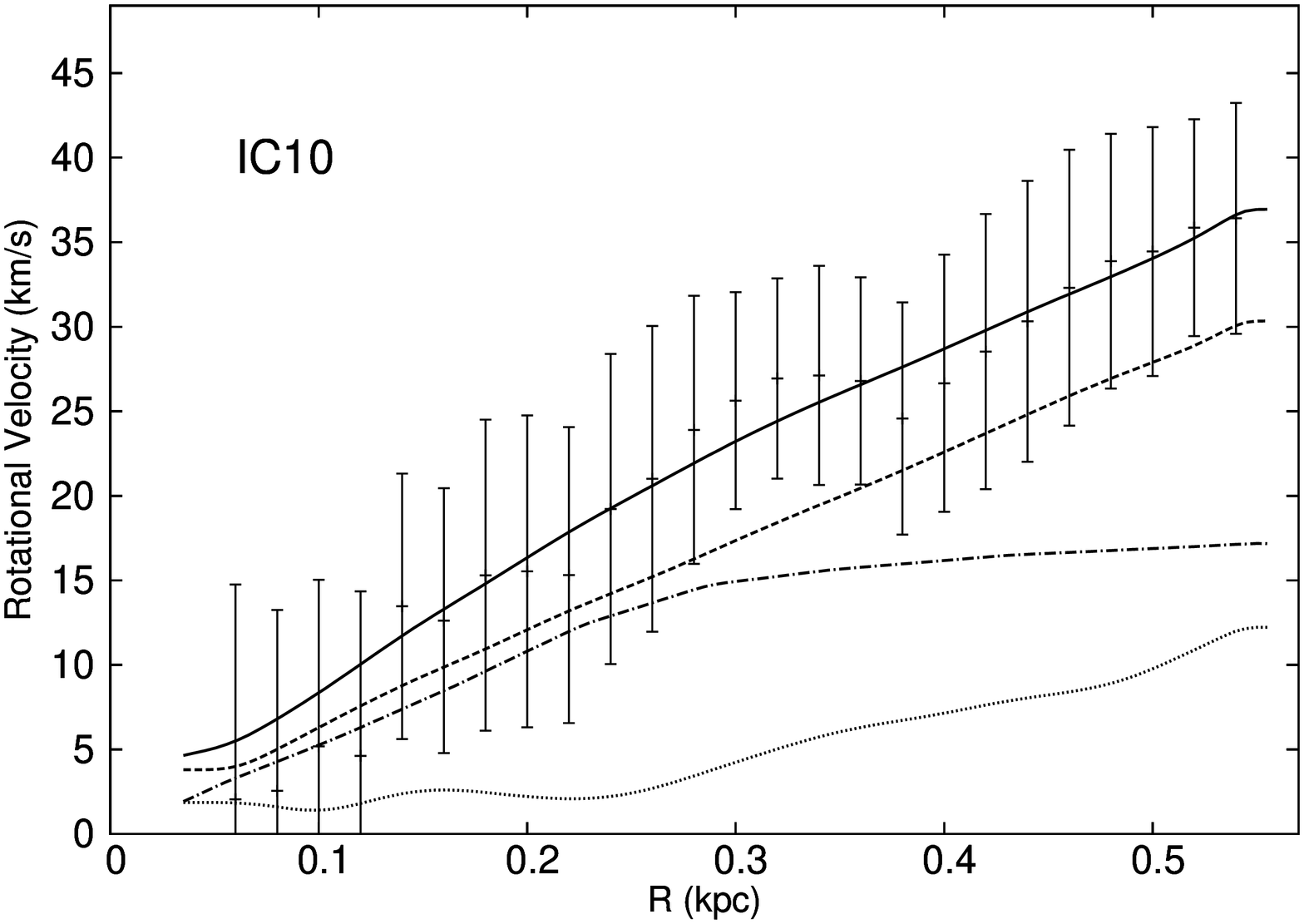,angle=270,width=7.05cm}}
\vskip 0.53cm
\hspace{0.11cm}
\leftline{\epsfig{file=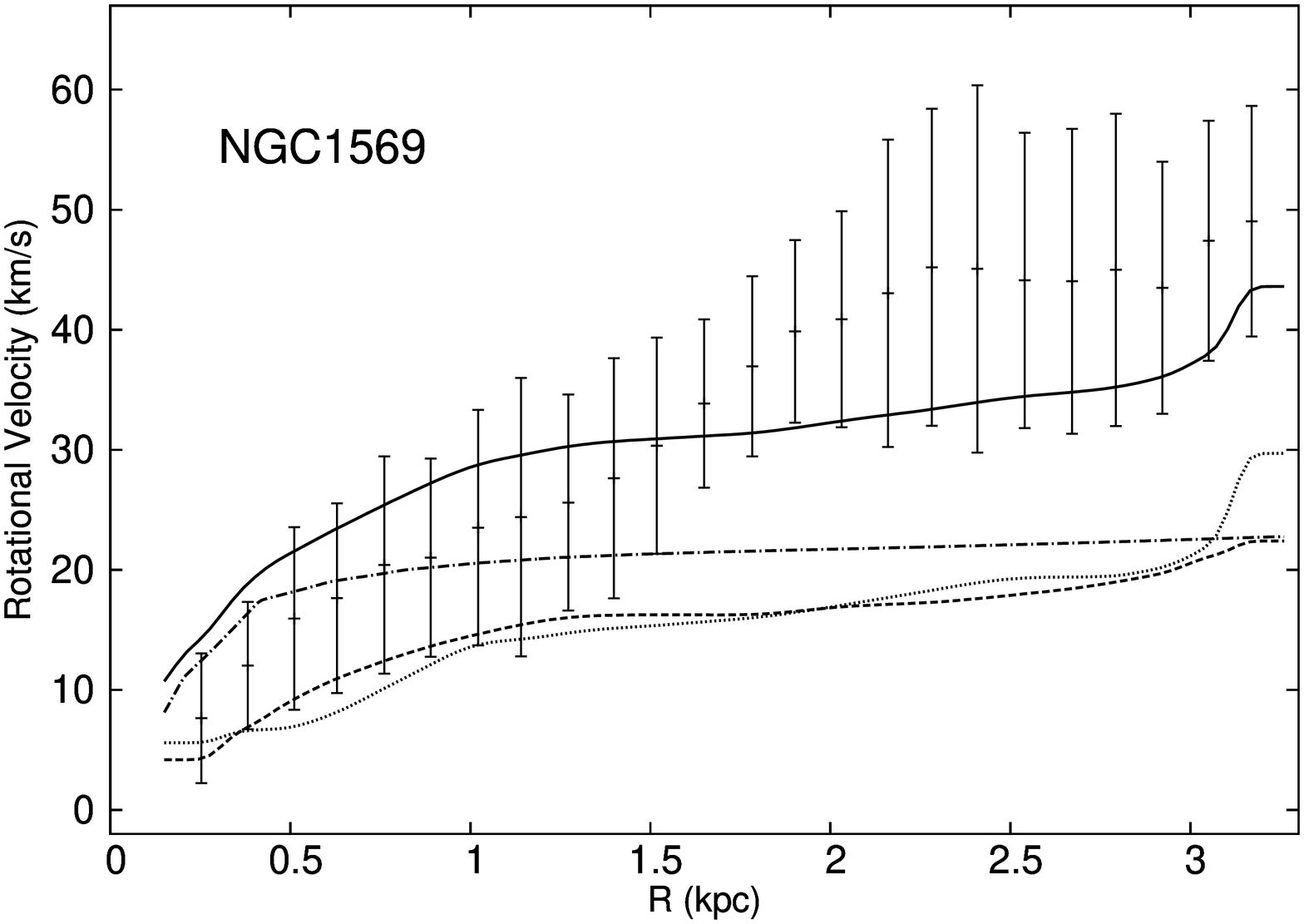,angle=270,width=7.05cm}}
\vskip -4.98cm
\hspace{-1.1cm}
\rightline{\epsfig{file=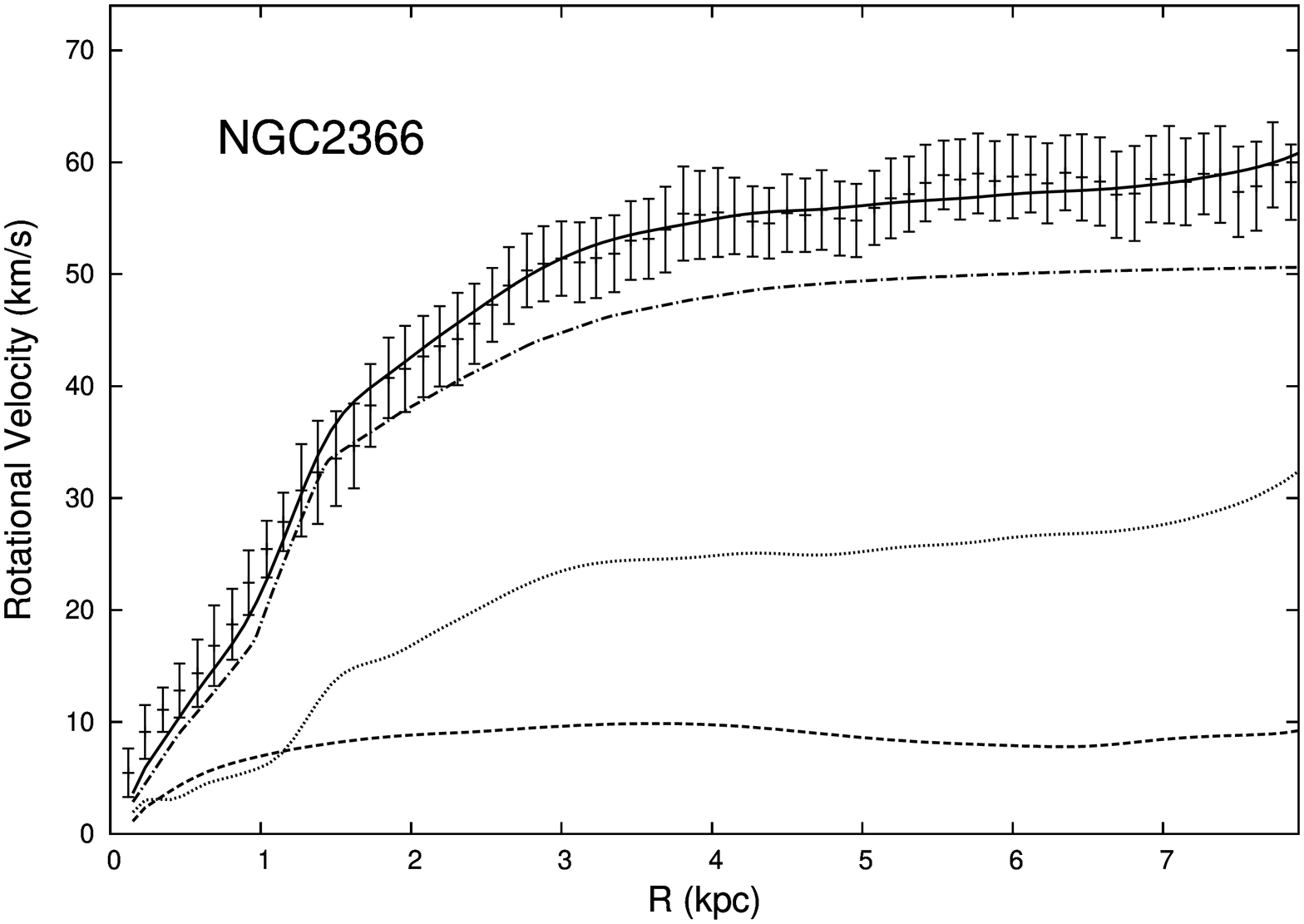,angle=270,width=7.05cm}}
\vskip 0.53cm
\hspace{0.11cm}
\leftline{\epsfig{file=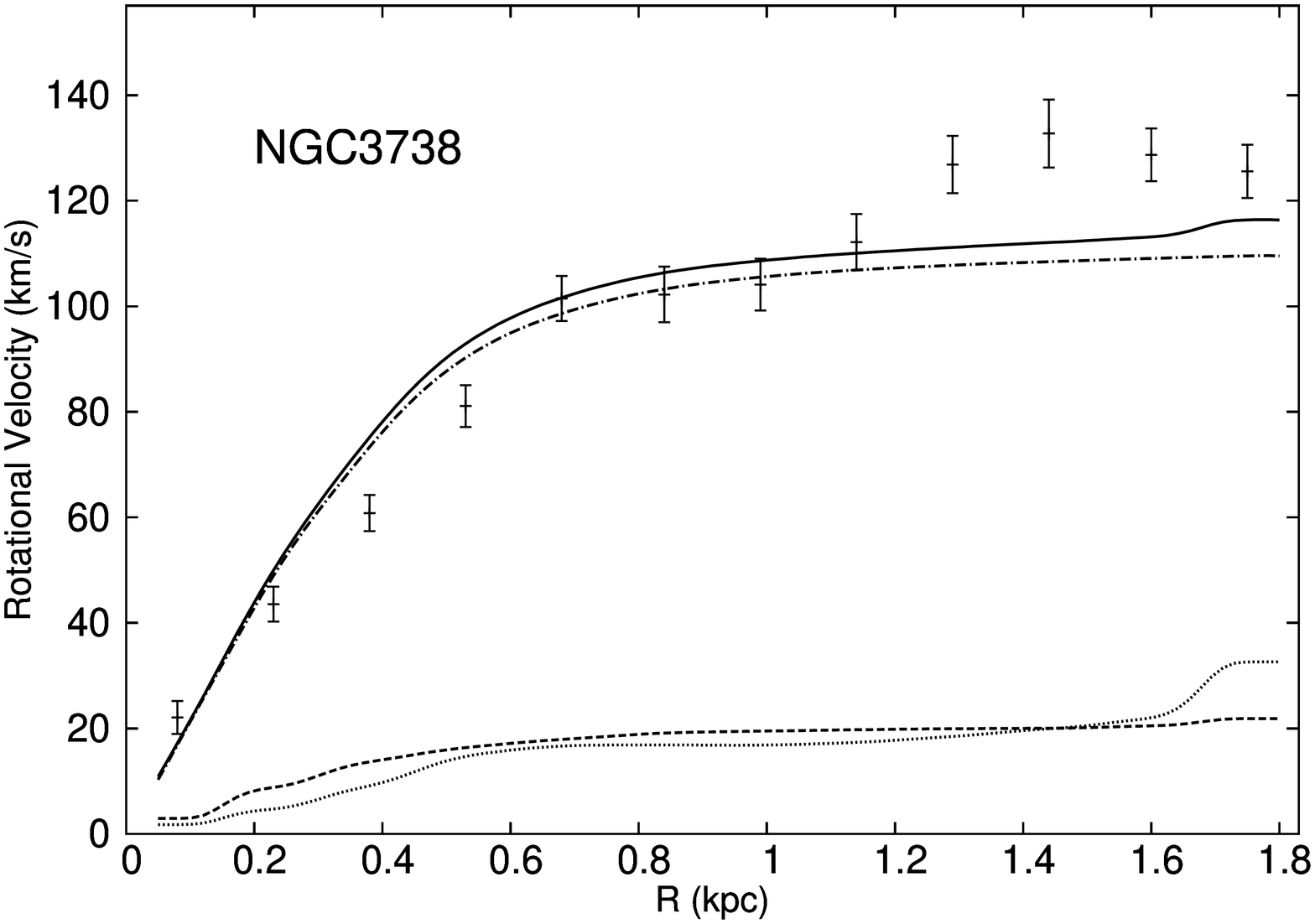,angle=270,width=7.05cm}}
\vskip -4.98cm
\hspace{-1.1cm}
\rightline{\epsfig{file=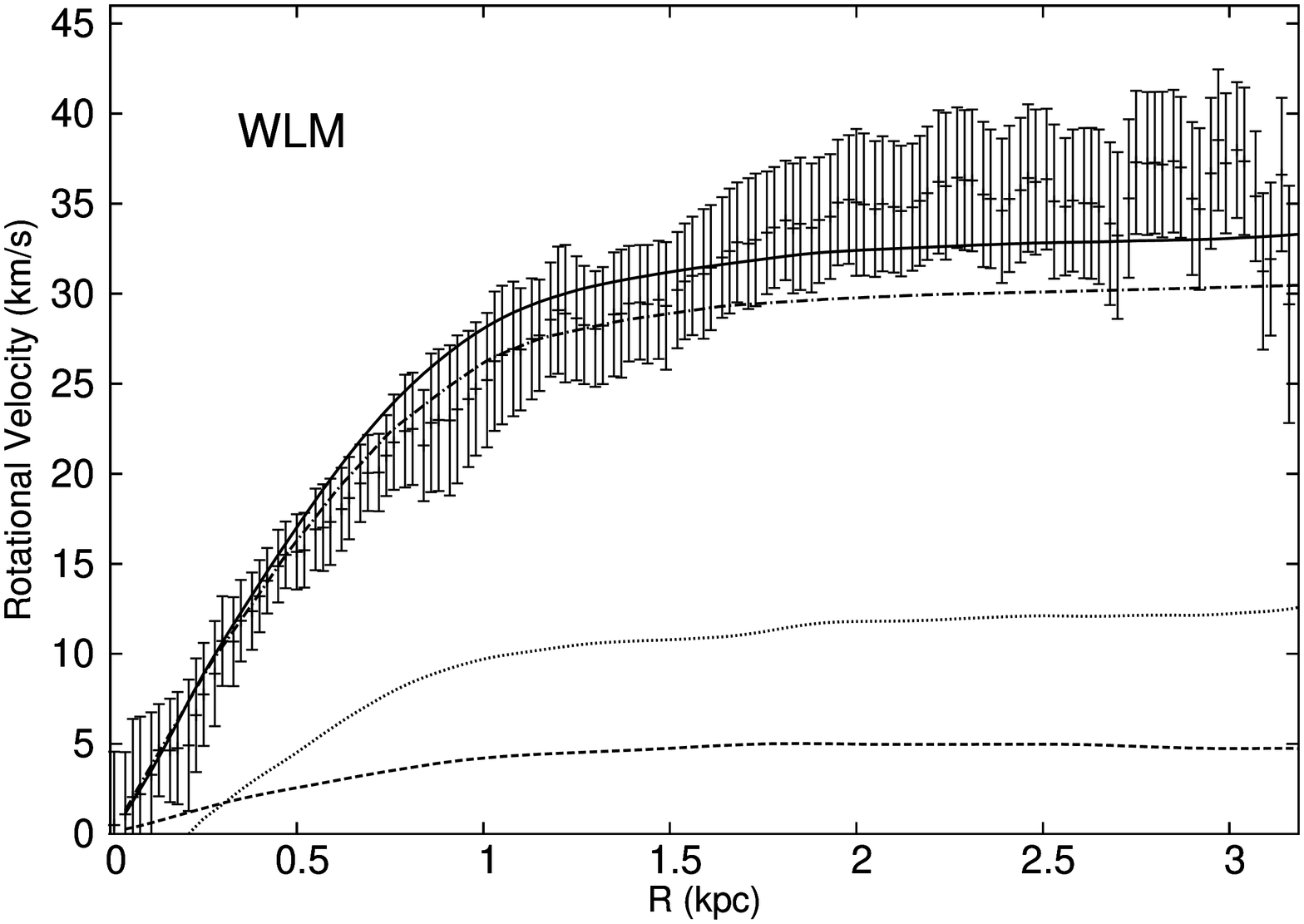,angle=270,width=7.05cm}}
\vskip 0.51cm
\noindent
{\small Figure 3: Rotation curves for the dwarf galaxies, DDO154, DDO210, DDO216, IC10, NGC1569,
NGC2366, NGC3738 and WLM.
Notation as in figure 2.
}

\vskip -1.0cm
\hspace{0.11cm}
%\centerline{\epsfig{file=fig1.eps,angle=0,width=11.0cm}}
\leftline{\epsfig{file=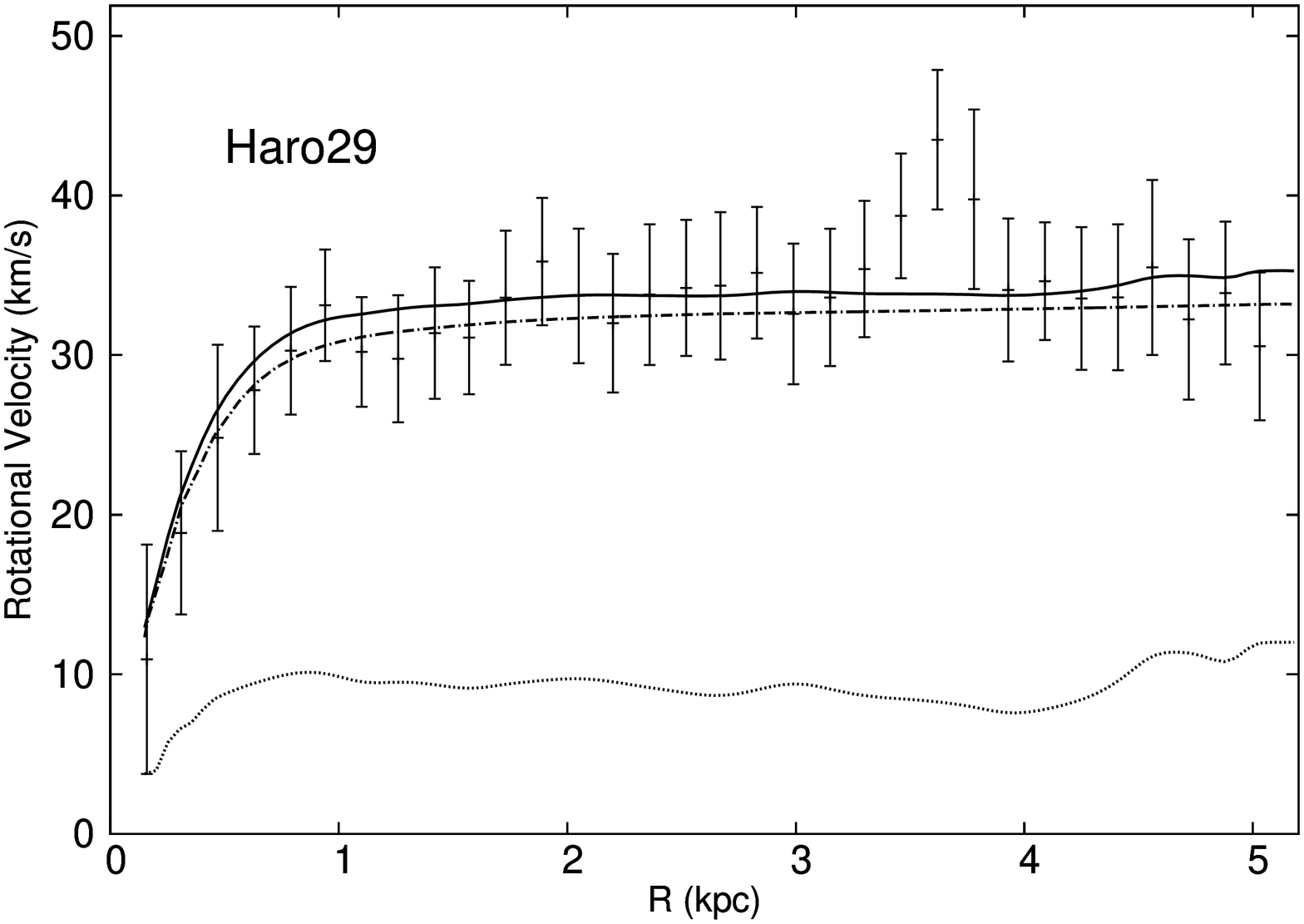,angle=270,width=7.05cm}}
\vskip -4.98cm
\hspace{-1.1cm}
\rightline{\epsfig{file=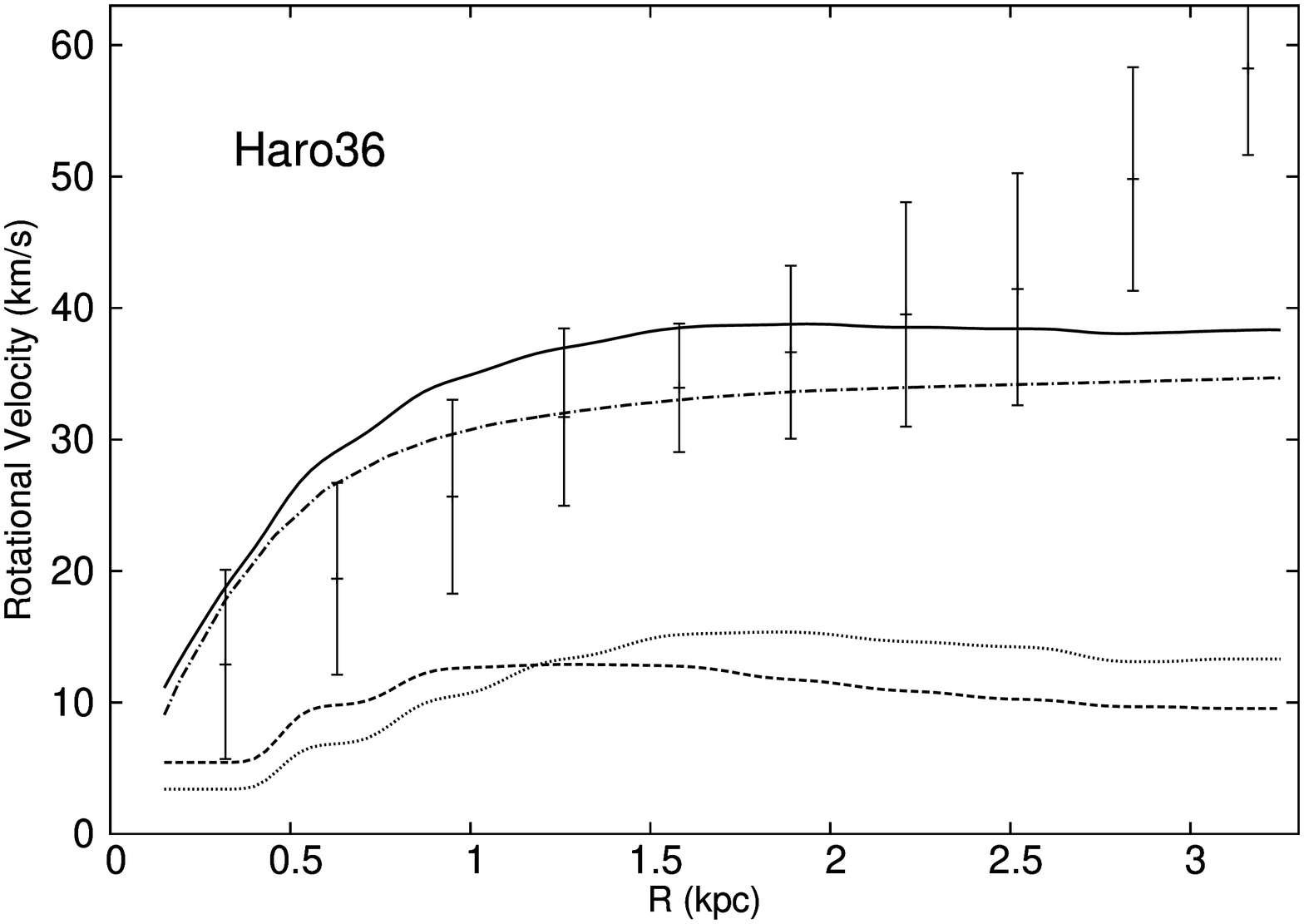,angle=270,width=7.05cm}}
\vskip 0.19cm
\hspace{0.11cm}
\leftline{\epsfig{file=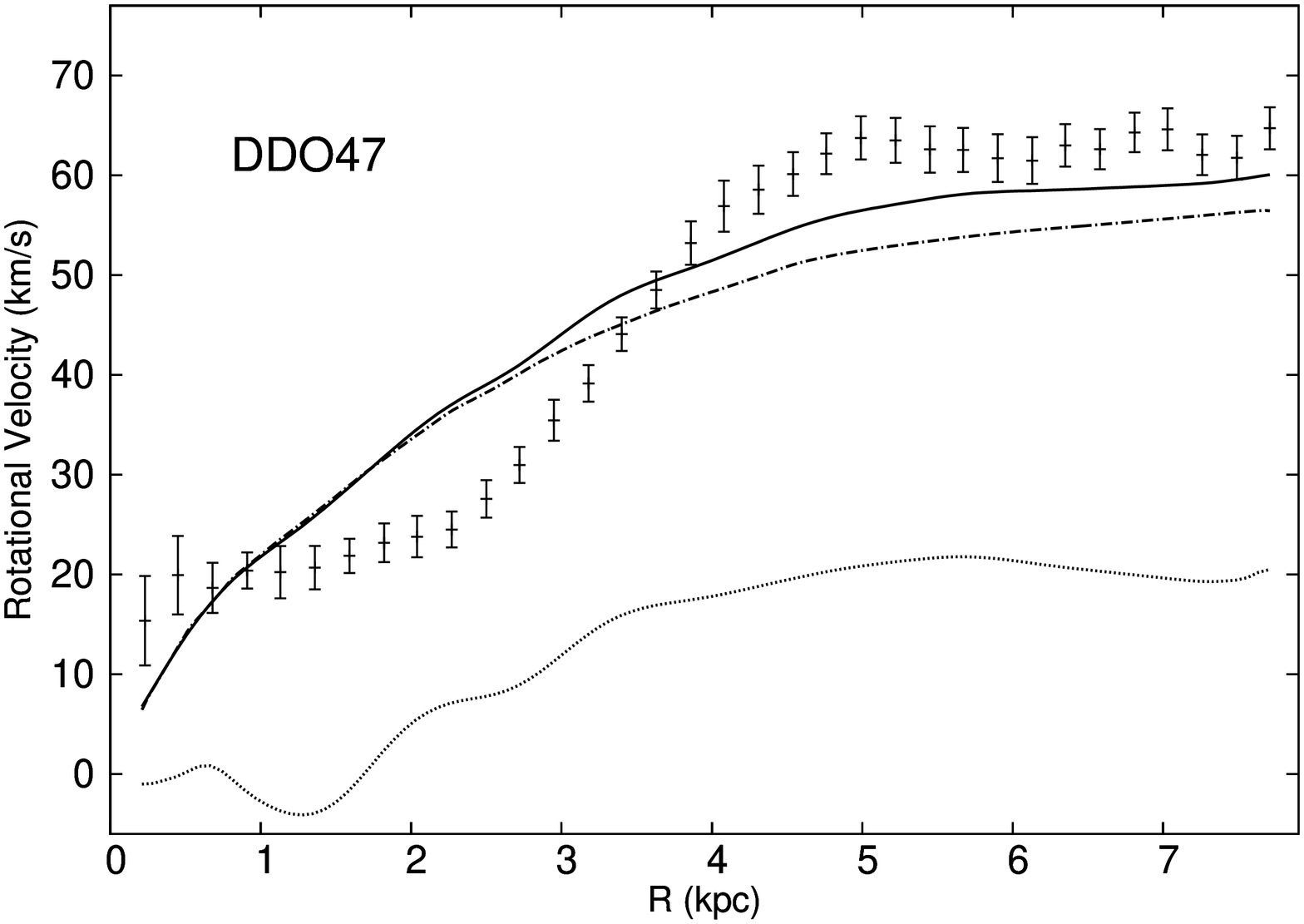,angle=270,width=7.05cm}}
\vskip 0.52cm
\noindent
{\small Figure 4: Rotation curves for the remaining two `classically' shaped dwarfs: Haro29 and Haro 36 and
also the irregularly shaped rotation curve of DDO47. Notation as in figure 2.}

\vskip 1.1cm
\hspace{0.11cm}
\leftline{\epsfig{file=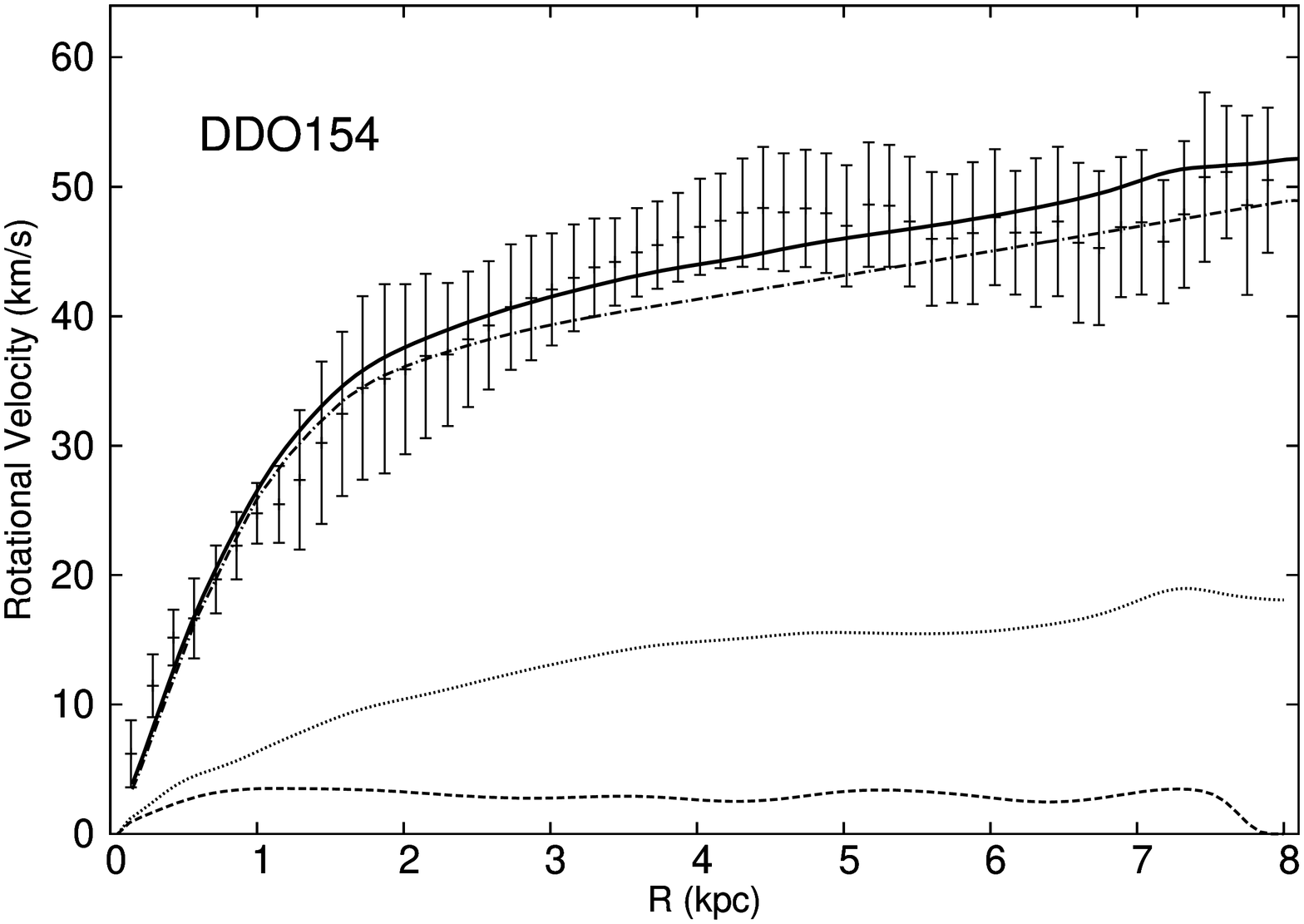,angle=270,width=7.05cm}}
\vskip -4.98cm
\hspace{-1.1cm}
\rightline{\epsfig{file=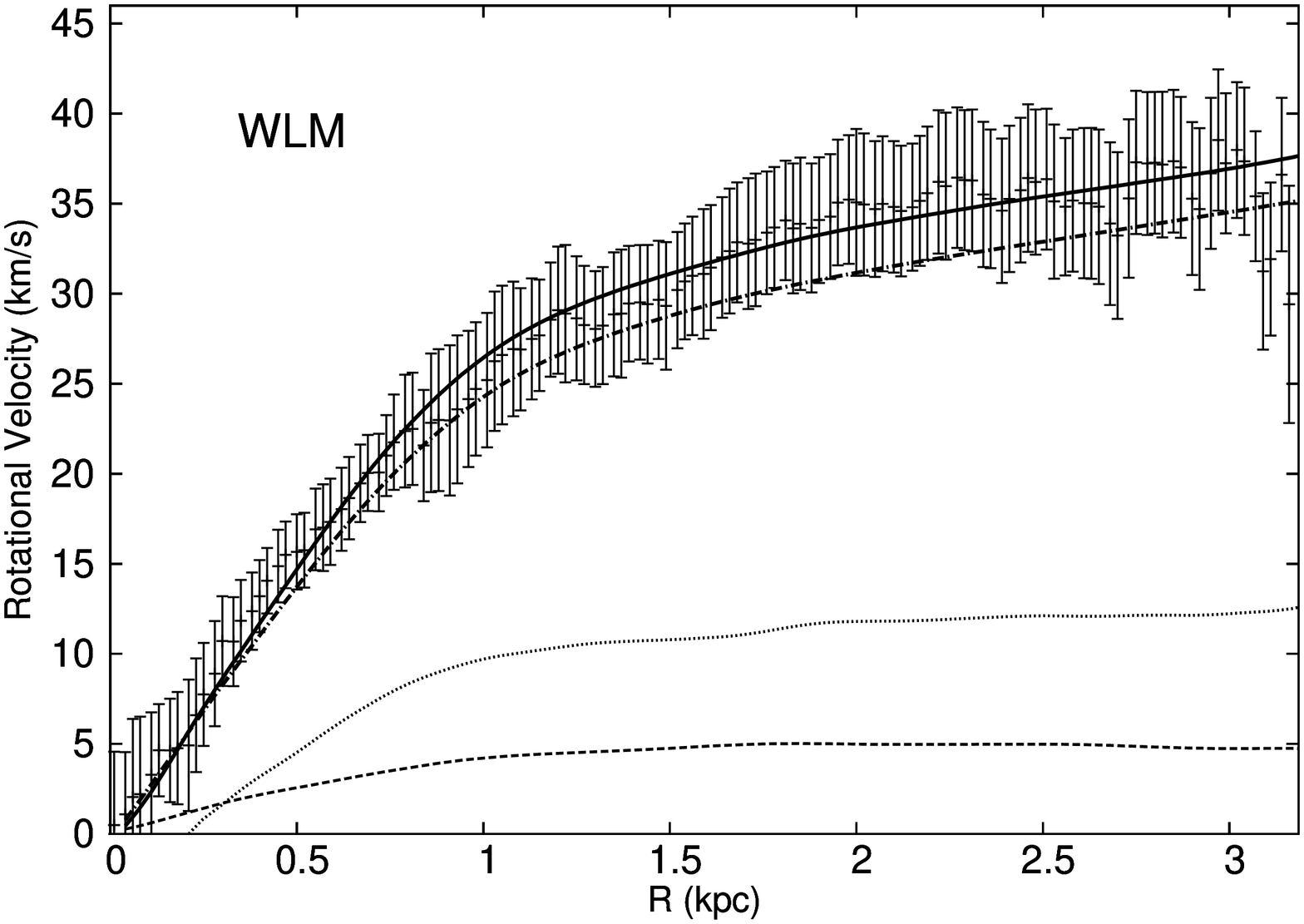,angle=270,width=7.05cm}}
\vskip 0.66cm
\noindent
{\small Figure 5: Rotation curves for DDO154 and WLM.
Notation as in figure 2, except that the
halo contribution was calculated from Eq.(\ref{3z}) allowing for a spatially dependent $\lambda$:
$\lambda(r) = \lambda_0 + \lambda_1 r$ and thus a two-parameter fit.
}

\newpage
%\vskip 0.35cm
\hspace{0.11cm}
%\centerline{\epsfig{file=fig1.eps,angle=0,width=11.0cm}}
\leftline{\epsfig{file=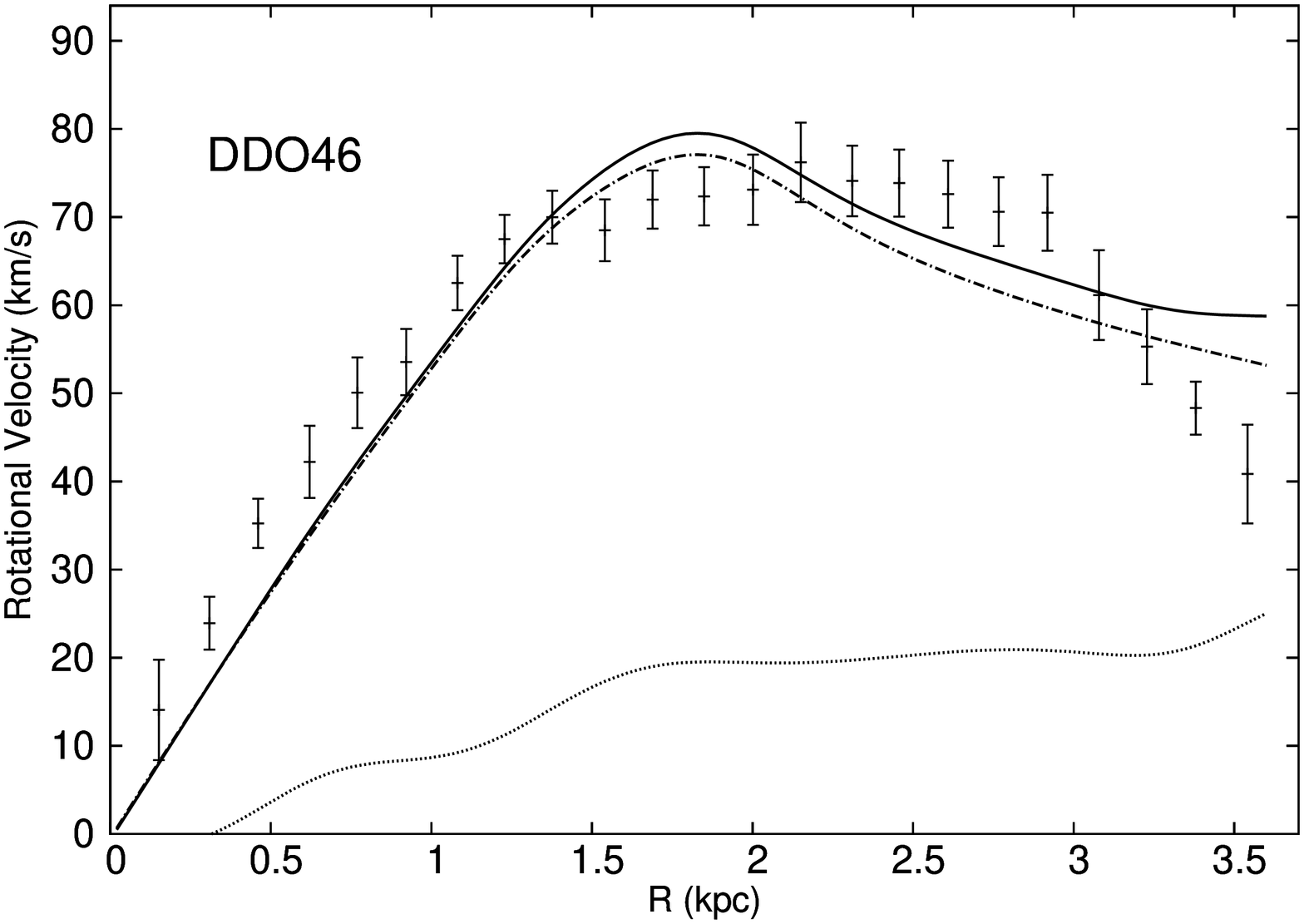,angle=270,width=7.05cm}}
\vskip -4.98cm
\hspace{-1.1cm}
\rightline{\epsfig{file=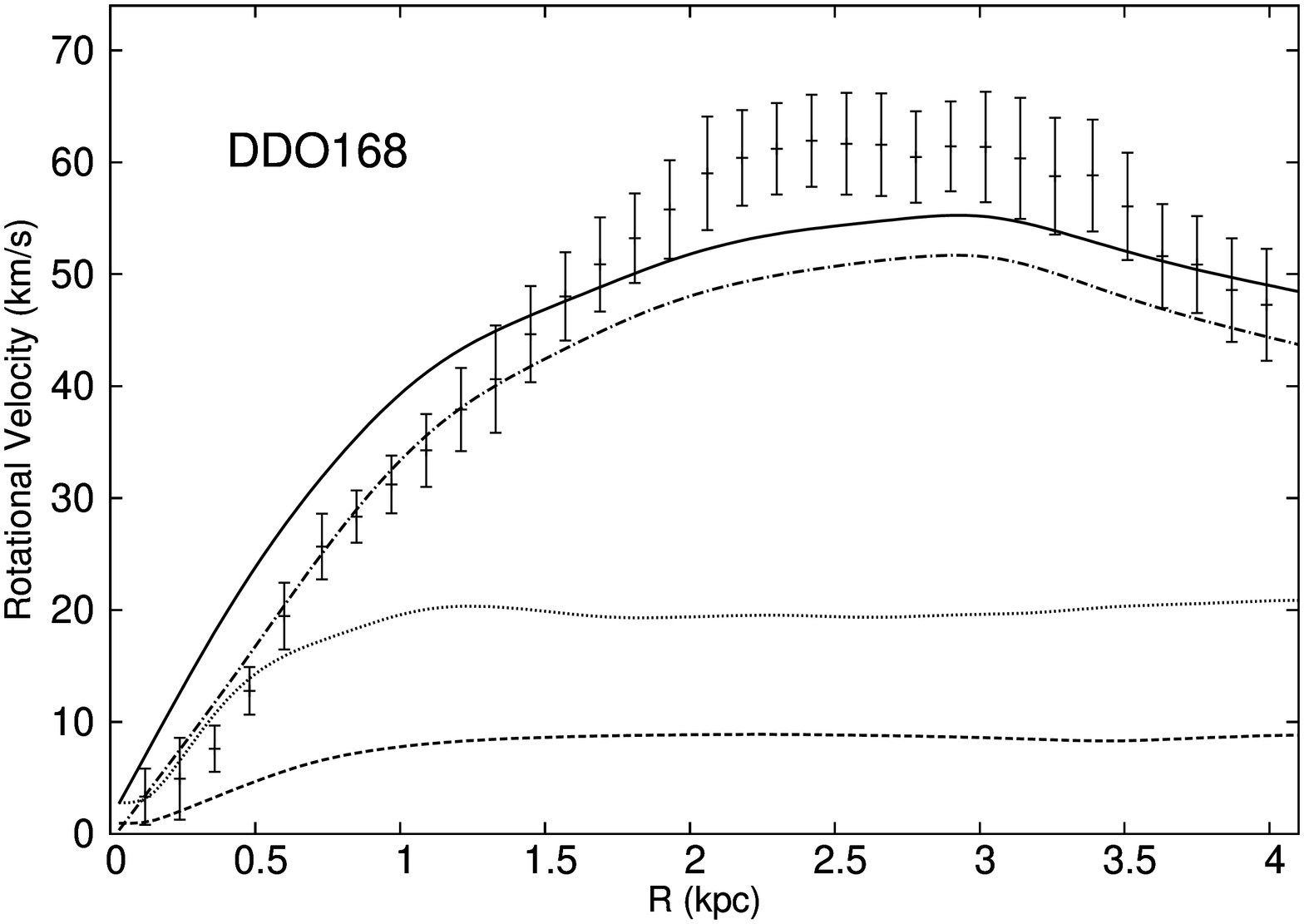,angle=270,width=7.05cm}}
\vskip 0.19cm
\hspace{0.11cm}
\leftline{\epsfig{file=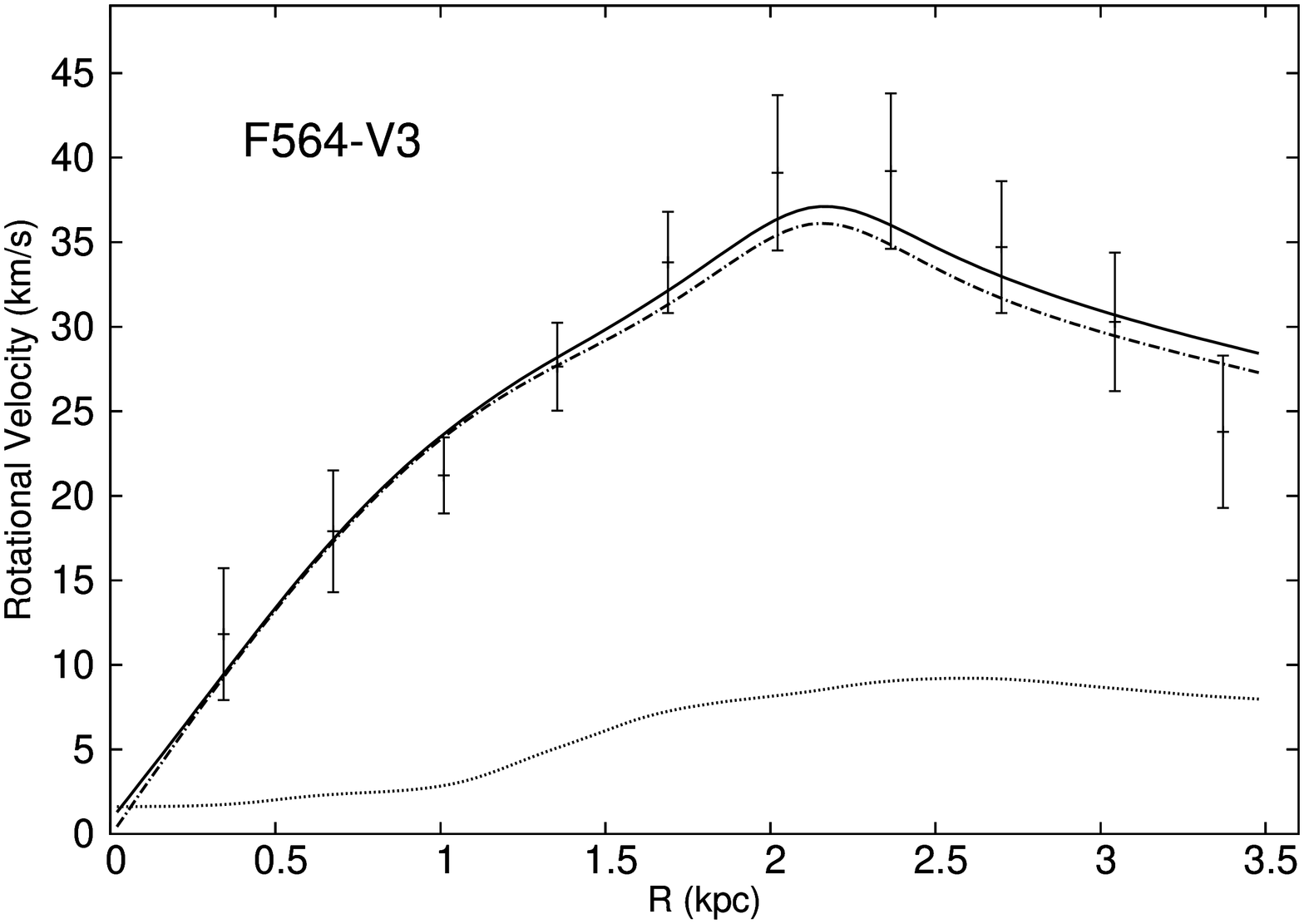,angle=270,width=7.05cm}}
\vskip 0.27cm
\noindent
{\small Figure 6: Rotation curves for the three dwarf galaxies, DDO46, DDO168, F564-V3, indicating a downturn in $v_{rot}$.
The downturn is modelled with a hardcut off in the dark matter density, as described in the text.
Notation as in figure 2.}

\vskip 0.4cm
\hspace{0.11cm}
%\centerline{\epsfig{file=fig1.eps,angle=0,width=11.0cm}}
\leftline{\epsfig{file=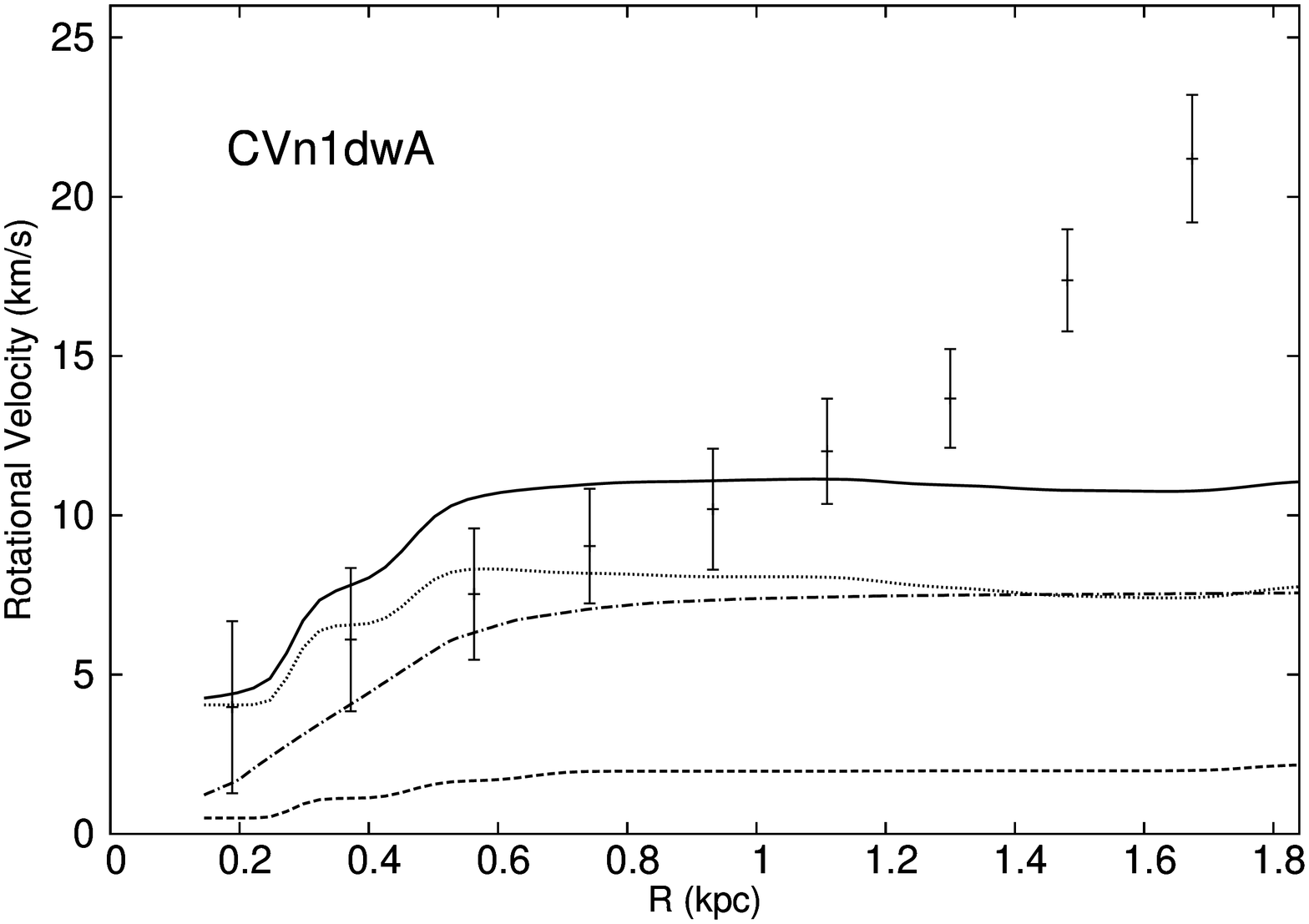,angle=270,width=7.05cm}}
\vskip -4.98cm
\hspace{-1.1cm}
\rightline{\epsfig{file=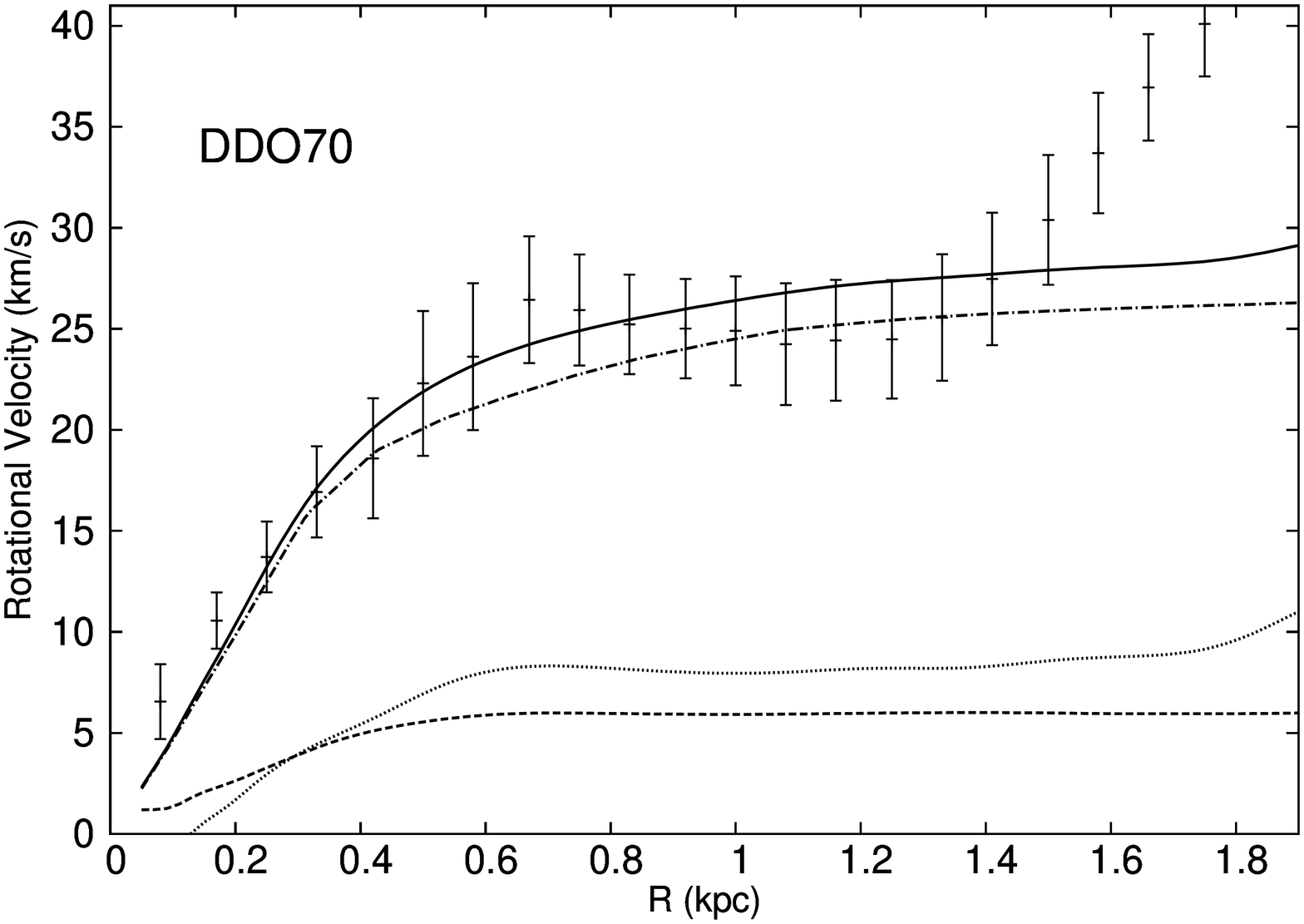,angle=270,width=7.05cm}}
\vskip 0.19cm
\hspace{0.11cm}
\leftline{\epsfig{file=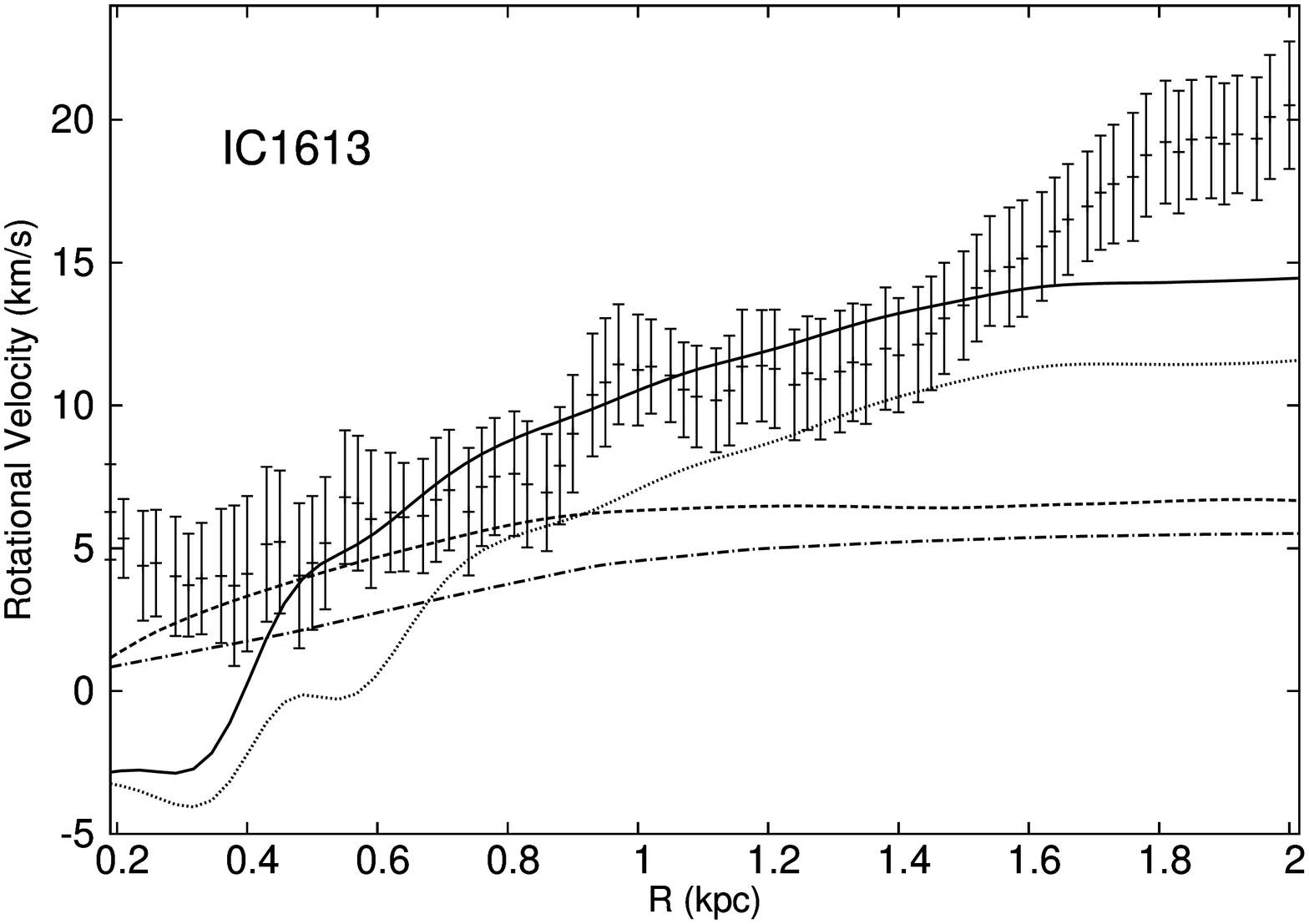,angle=270,width=7.05cm}}
\vskip -4.98cm
\hspace{-1.1cm}
\rightline{\epsfig{file=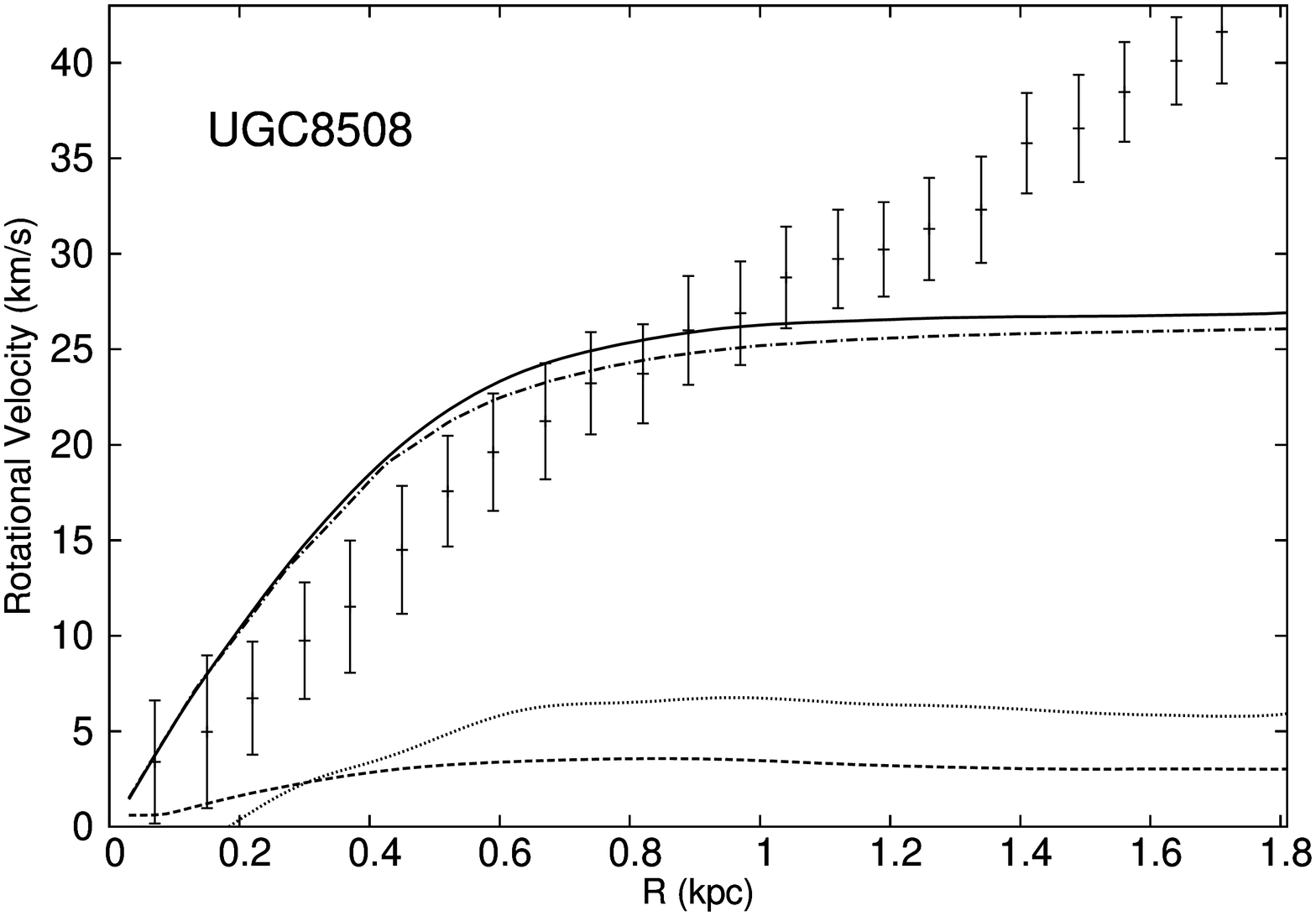,angle=270,width=7.05cm}}
\vskip 0.53cm
\noindent
{\small Figure 7: Rotation curves for the remaining LITTLE THINGS dwarfs [CVn1dwA, DDO70, IC1613, 
UGC8508] 
which feature a `hump' at the largest measured radii. 
Shown are fits to the inner part of the rotation curve data, see discussion in text.  
Notation as in figure 2.}

%\newpage

%%%%%%%%%%%% paragraph about things spirals %%%%%%
\vskip 1.2cm

\subsection{THINGS spirals}
\vskip 0.2cm

Let us now briefly comment on spiral galaxies.
%We have also examined spiral galaxies from the THINGS survey \cite{things}.
Compared to dwarf irregular galaxies,
spirals have a much larger stellar component, which typically dominates over the dark matter contribution in the inner part
of the galaxy. This stellar component is poorly constrained due to uncertain stellar mass-to-light ratio, $\Upsilon_*$. 
This unfortunately tends to make spirals much less useful in testing dark matter models.
Also, the UV and $H\alpha$ flux measurements for spirals have larger uncertainties due to much larger extinction corrections.
A  sample of spirals with high resolution rotation curves has been given by the THINGS collaboration \cite{things}.
An example is given in figure 8, 
where the fit to the rotation curve of NGC2403 is given. In this example the baryonic contribution was obtained from \cite{things} and 
modelled allowing $\Upsilon_*$ to vary, and the dark matter contribution was obtained from Eq.(\ref{4z}) and Eq.(\ref{3z}). 
The supernovae rate in Eq.(\ref{3z})
was modelled via a Kennicutt-Schmidt relation, with exponent $N=2$, $\Sigma_{SN} \propto \Sigma_{gas}^2$
(UV surface profile was unavailable). 

\begin{table}
\centering
\begin{tabular}{c c c  c  }
\hline\hline
Galaxy & D (Mpc) &  $M_{FUV}$ & $\stackrel{\sim}{\lambda}$  
(km$^2$/s$^2$) 
{\rule{0pt}{2.9ex}}       
{\rule[-1.5ex]{0pt}{0pt}} 
\\
\hline
NGC925 & 9.2 & -18.28 & \ 6.1E-4
{\rule{0pt}{2.9ex}}       
\\
NGC2403 & 3.2 & -17.85 & 1.2E-3
\\
NGC2841 & 14.1 & -18.77 & 2.0E-3
\\
NGC2903 & 8.9 & -18.45 & 1.2E-3
\\
NGC2976 & 3.6 & -14.67 & 6.6E-3
\\
NGC3031 & 3.6 & -18.08 & 4.8E-4
\\
NGC3198 & 13.8 & -18.97  & 4.4E-4
\\
NGC3521 & 10.7 & -18.37 & 1.5E-3
\\
NGC3621 & 6.6 & -18.29 & 8.8E-4
\\
NGC4736 & 4.7 & -16.86 & 6.5E-4 
\\
NGC5055 & 10.1 & -18.57 & 9.4E-4
\\
NGC6946 & 5.9 & -18.74 & 8.4E-4
\\
NGC7331 & 14.7 & -18.96 & 7.5E-4
\\
NGC7793 & 3.9 & -17.18 & 1.5E-3
{\rule[-1.5ex]{0pt}{0pt}} 
\\
\hline\hline
\end{tabular}
\vskip 0.3cm
\caption{
{\small
THINGS spirals:
$D$ is the distance of the spiral and $M_{FUV}$ 
is the FUV absolute AB magnitude corrected for internal and foreground extinction.  
Also given is the parameter $\stackrel{\sim}{\lambda}$, Eq.(\ref{10x}).
}
}
\end{table}

\vskip 0.7cm
\centerline{\epsfig{file=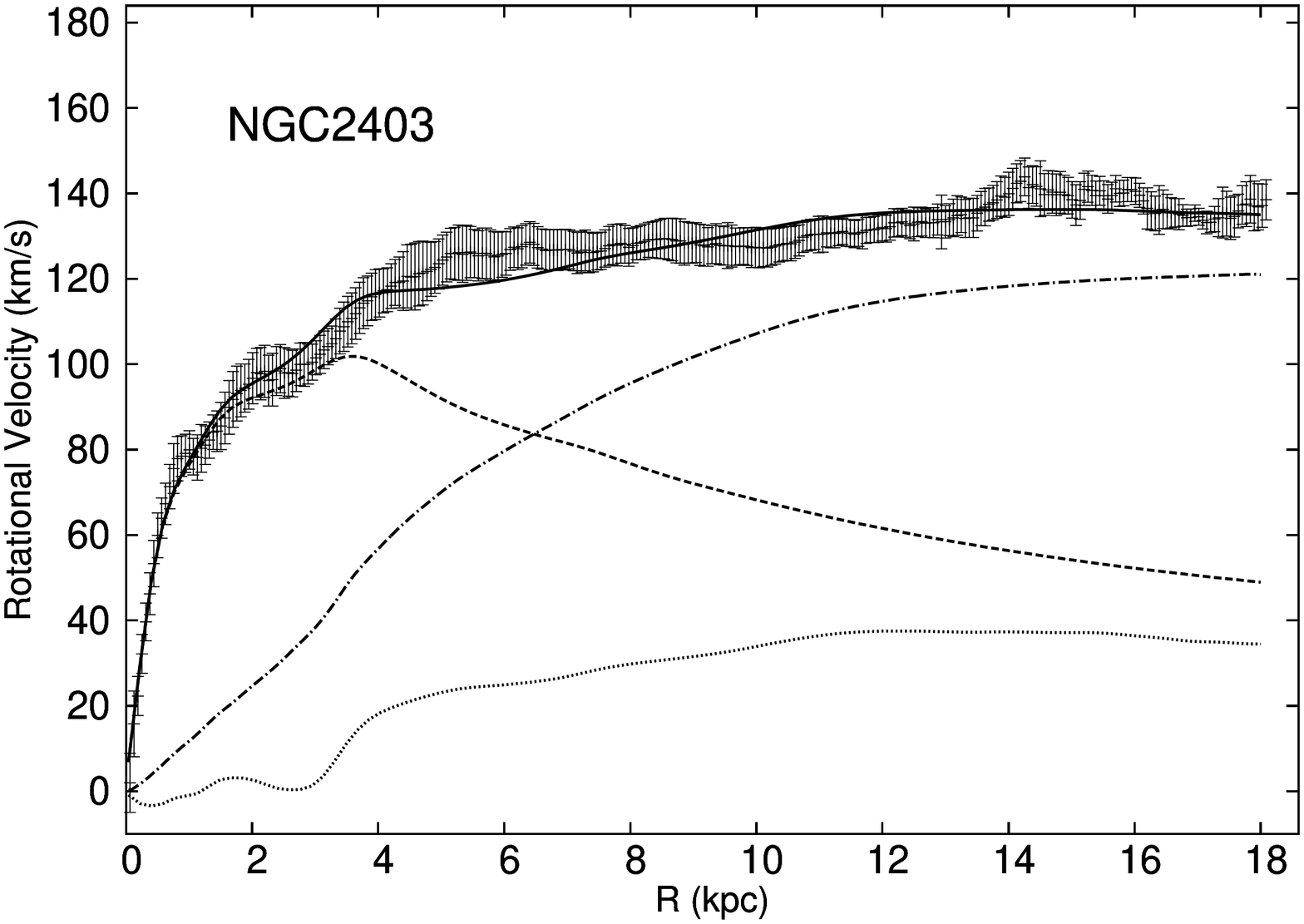,angle=270,width=11.3cm}}
\vskip 0.2cm
\noindent
{\small Figure 8: The spiral galaxy, NGC2403. 
The stellar (dashed line), baryonic gas (dotted line) and dissipative dark halo (dashed dotted line) contributions
are shown. The solid line is the sum of these contributions added in quadrature.
}

%%%%%%%%%%%%%%%%%%
\subsection{The scaling of $\lambda$}

In the fits of the dwarf galaxies the parameter $\lambda$ in Eq.(\ref{3z}) was allowed to have a different value
for each galaxy. In specific dissipative dark matter models this parameter is expected to have a
more universal character, with
simple arguments suggesting a weak scaling: $\lambda \propto 1/v_{rot}^{max}$
[see earlier discussion around Eq.(\ref{T})].

At sufficiently large radii
the dark photon flux falls at the geometric rate: $F_{\gamma_D} \propto 1/r^2$.
At such large radii the dark matter halo mass density profile 
[Eq.(\ref{3z})] reduces to:
\begin{eqnarray}
\rho (r) = \lambda {R_{SN}
\over 4\pi r^2}
\ . 
\end{eqnarray}
Here $R_{SN}$ is the integrated supernova rate in the disk.
Eq.(\ref{4z}) gives
an asymptotically flat value for the halo contribution to the circular velocity:
\begin{eqnarray}
v_{halo}^2 = \lambda G_N R_{SN}
\ .
\label{s5x}
\end{eqnarray}
Supernovae are the final evolutionary stage of large stars. As discussed earlier, the FUV luminosity ($L_{FUV}$) is expected to provide a
reasonably 
direct estimate of the 
star formation rate, so that:
$R_{SN} \propto L_{FUV} \propto 10^{-0.4 M_{FUV}}$. It is convenient then to introduce the quantity $\stackrel{\sim}{\lambda}$: 
\begin{eqnarray}
\stackrel{\sim}{\lambda} \ \equiv \ {v_{halo}^2 \over 10^{-0.4 M_{FUV}}}
\label{10x}
\ .
\end{eqnarray}
With this definition Eq.(\ref{s5x})
implies $\stackrel{\sim}{\lambda} = \lambda G_N c_1$ where
$c_1 \equiv R_{SN}/10^{-0.4 M_{FUV}}$.
For each galaxy, $v_{halo}$ 
can be obtained from the fit of the model to the rotation curve data (for the dwarfs with a downturn in $v_{rot}$, listed in table 2, we set
$v_{halo} = v_{rot}^{max}$). The absolute FUV AB magnitude, $M_{FUV}$, has already been provided  
in tables 1-3 for the 24 dwarfs for which FUV flux measurements are available. 
These tables also give the 
resulting $\stackrel{\sim}{\lambda}$ values, as defined in Eq.(\ref{10x}), for each dwarf.
%In figure 9 we plot $\stackrel{\sim}{\lambda}$ versus $v_{rot}^{max}$ for these 24 dwarfs.
We have also estimated the 
$\stackrel{\sim}{\lambda}$ values for 
the THINGS spirals. 
For these spirals we obtained  $v_{halo}$ from the quasi-isothermal fit with free mass-to-light stellar mass
parameter as given in \cite{things}. 
The absolute FUV magnitude was obtained using data from \cite{table,tablened} and corrected for both internal and foreground extinction
following the procedure of \cite{flux,flux2}.
The results of this exercise are given in table 4.  
Figure 9 summarizes these results.\footnote{
The uncertainty in the parameter $\stackrel{\sim}{\lambda}$ 
was estimated from the distance uncertainty, $v_{halo}$ uncertainty, and the uncertainty in the extinction corrections.
The distance uncertainty for most of the galaxies is small, typically $10\%$, while the $v_{halo}$ uncertainty was taken from the rotation
velocity data points  
at large radii.
The uncertainty in extinction was assumed to be $\pm 20\%$ of the extinction correction.
}

Under the assumptions that the halo heating is approximately independent
of the halo temperature, $T$,
and halo cooling is dominated by bremsstrahlung, $\Gamma_{cool} \propto \sqrt{T} n^2$, one expects
the rough scaling
$\lambda \ \propto 1/\sqrt{T}$. 
As discussed earlier in section 2 [see discussion around Eq.(\ref{T})],
this leads to the correlation:
$\lambda  \ \propto 1/v_{rot}^{max}$. 
Assuming that $c_1$ (defined above) is constant this then implies
$\stackrel{\sim}{\lambda} \ \propto 1/v_{rot}^{max}$ or equivalently,
\begin{eqnarray}
L_{FUV} \propto v_{halo}^2 v_{rot}^{max}
\ .
\label{bom}
\end{eqnarray}
This scaling is also shown in figure 9.
For many galaxies one could make the rough approximation:  $v_{halo} \approx v_{rot}^{max}$, from which Eq.(\ref{bom}) reduces to:
$L_{FUV} \propto [v_{rot}^{max}]^3$.
For spiral galaxies,
this is approximately equivalent to the Tully-Fisher relation \cite{tf}, e.g. the study \cite{webster} found
that $L_B \propto [v_{rot}^{max}]^\alpha$, where $\alpha = 3.4 \pm 0.09$ (the difference between this exponent and the value
$\alpha = 3$ is consistent with the slight scaling difference between the FUV and B-bands).

Figure 9 shows a significant scatter of the $\stackrel{\sim}{\lambda}$ parameter for the LITTLE THINGS dwarf galaxies.
Some scatter can be due to uncertainties, e.g. the determination of the galaxy's absolute magnitude
will be affected by 
the uncertainty in its distance, the parameter $c_1$ might vary somewhat between different galaxies etc., 
however such uncertainties are unlikely to be the whole explanation \footnote{
Note  however that since
Eq.(\ref{3z}) defines the dark matter density directly in terms of the supernova density (rather than say, baryonic mass), 
the estimated values for $\stackrel{\sim}{\lambda}$ should be independent of the uncertain mass-to-light
ratio of the galaxy. 
}.
The size of the scatter might indicate fundamental physical differences between galaxies, or it could simply be an indication
that the dissipative halos are not all in a steady-state configuration. Halo heating and cooling rates may be unbalanced due to
perturbations from nearby galaxies, current (or recent) starburst activity etc.
As already discussed, such activity might cause distortions including, possibly, the `bump' and `hump' in the rotation curves apparent in a
subset of the dwarfs.
If so, then 
it could be a factor behind the poor fits to the shape of the rotation curve that some of the dwarfs feature.
In figure 10 we therefore consider only a subset of galaxies:
the THINGS spirals together with the LITTLE THINGS classical dwarfs and
only those with $\chi^2_r (FUV) < 2$. This means that
the dwarfs DDO101, DDO50, DDO133, NGC3738  (and those of tables 2 and 3) are excluded, and
we also exclude NGC1569 which is undergoing extreme starburst activity \cite{starburst} 
(whose halo is therefore unlikely to be in an equilibrium configuration).

Figure 10 shows a much reduced  scatter (cf. figure 9) of the $\stackrel{\sim}{\lambda}$ values with respect to the 
$\stackrel{\sim}{\lambda} \ \propto 1/v_{rot}^{max}$ scaling. 
This is some evidence in favour of a `universal'  $\lambda$ parameter, presumably set in part by the fundamental particle
physics of the dissipative model.
Further studies, with a larger sample 
would be a useful next step. 
In particular, the possibile validity of Eq.(\ref{bom}) over the entire range from dwarfs to spirals warrants  further investigation,
along with possible connection with the baryonic Tully-Fisher relation \cite{btf}.

\vskip 0.7cm
\centerline{\epsfig{file=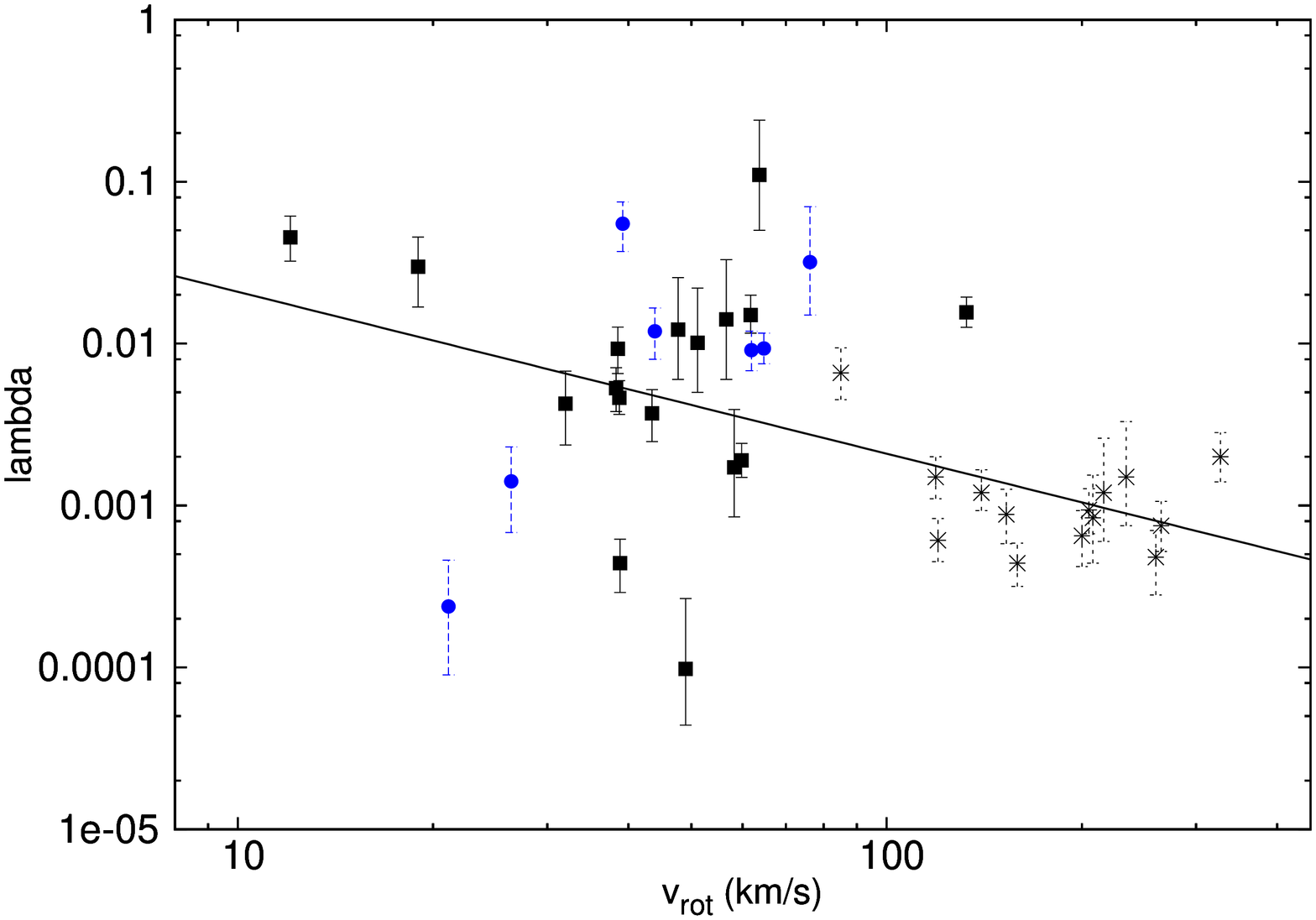,angle=270,width=11.7cm}}
\vskip 0.3cm
\noindent
{\small Figure 9: The parameter $\stackrel{\sim}{\lambda}$ [Eq.(\ref{10x})] versus $v_{rot}^{max}$ for the LITTLE THINGS dwarfs (squares and
circles) and THINGS
spirals (stars).  
The squares are the classical dwarfs while the circles are all the others.
The solid line is the 
$\stackrel{\sim}{\lambda} \ \propto 1/v_{rot}^{max}$ scaling discussed in text.}
\vskip 1.1cm
\centerline{\epsfig{file=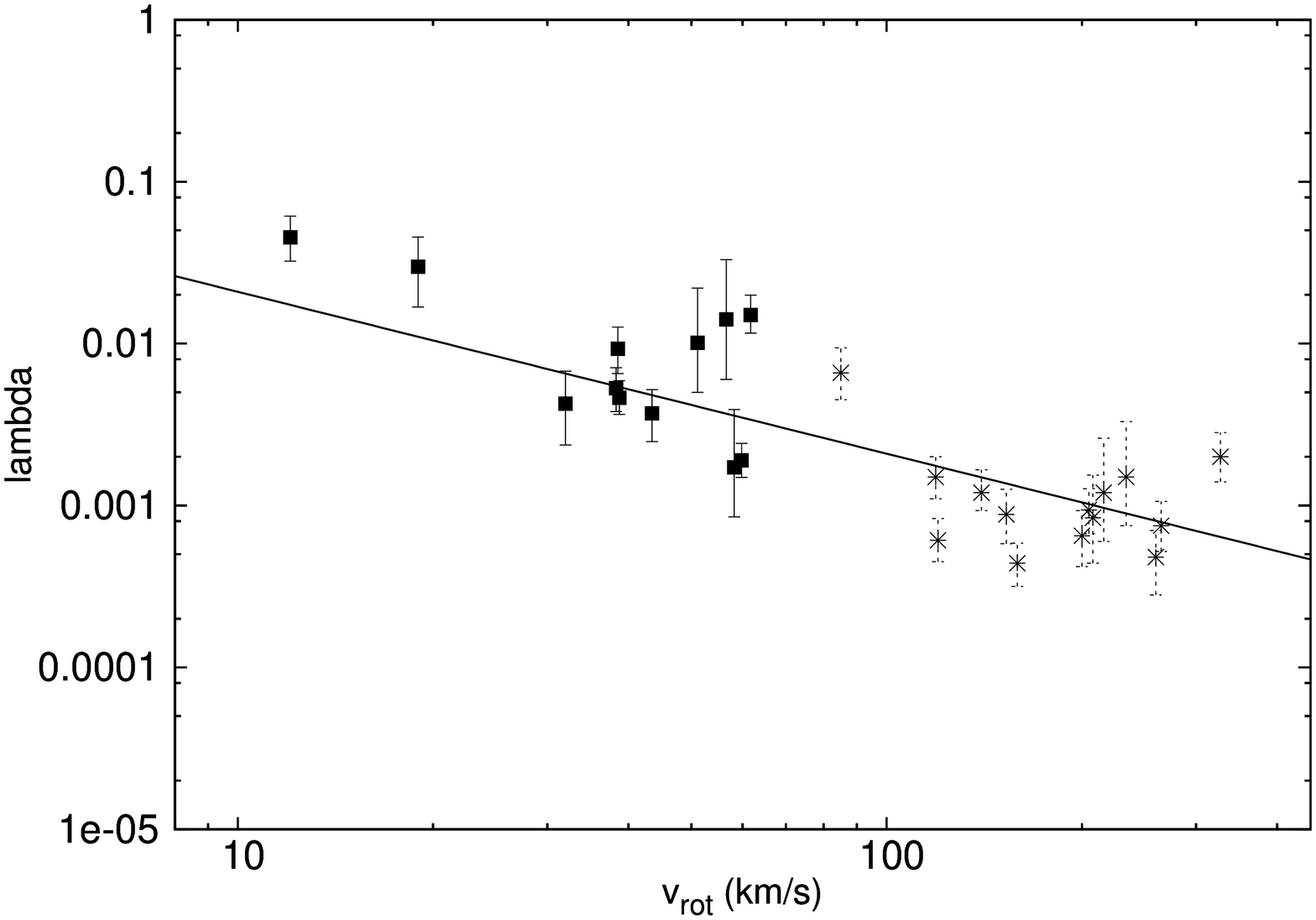,angle=270,width=12.0cm}}
\vskip 0.3cm
\noindent
{\small Figure 10: Same as figure 9 except only THINGS spirals and `best fit' classical dwarfs are included (see text).
}

\vskip 1.1cm

\vskip 0.3cm
\section{Conclusion}

There is substantial evidence from rotation curves that dark matter halos have nontrivial dynamics.
Most striking is that 
the observed rotation curves are strongly correlated with the baryon content.
The Tully-Fisher relation provides a global relation, while the dark matter 
rotational velocity generally transits from linear rise (near the galactic center) to flat at a distance scale associated
with the baryonic distribution, and there are also many other correlations.

Dark matter halos around disk galaxies can have such nontrivial dynamics if dark matter
is strongly self interacting and dissipative. 
Multicomponent hidden sector dark matter featuring a massless `dark photon' (from an unbroken dark $U(1)$ gauge
interaction) which kinetically mixes with the ordinary photon provides a concrete 
example of such dark matter. The kinetic mixing interaction facilitates halo heating by enabling
ordinary supernovae to be a source of these dark photons. 
Dark matter halos can expand and contract
in response to the heating and cooling processes, but for a sufficiently isolated and unperturbed halo
should have evolved to a steady state or `equilibrium' 
configuration where heating and cooling rates locally balance. 
This dynamics allows the current dark matter density profile to be related to the distribution of ordinary supernovae
in the disk of a given galaxy, Eq.(\ref{3z}). 
Naturally, this is a simplified description and would likely be  invalid for galaxies strongly influenced by perturbations, 
starburst activity etc. 

The ordinary supernovae distribution was here modelled via the UV light emission profile,
which should provide a fairly direct measurement of this distribution in a given galaxy.
This improves upon previous work of \cite{foot5} which used the gas density via a Kennicutt-Schmidt type relation.
The resulting halo rotation curve then depends on only one parameter $\lambda$ (assumed to be spatially independent as
a zeroth order approximation) and tested against the full LITTLE THINGS sample of 
26 dwarf galaxies.
The analysis indicates that for most of the LITTLE THINGS dwarfs, the dark matter 
predictions resulting from the simple formula, Eq.(\ref{3z}), agree
reasonably well with the observations. Not all of the dwarfs analysed have good agreement, however
the  poor fits of some of these dwarfs can plausibly be ascribed to environmental
influences which perturb the halo.
Such a perturbation may only be temporary, the timescale depending on the nature of the 
perturbation and the details of the dissipative particle physics.

In the simplified approach considered here,
the steady-state configuration, Eq.(\ref{3z}), is described by a single parameter, $\lambda$. 
This parameter is a function of various fundamental parameters, e.g. in the model of \cite{foot4} one has dark particle masses, dark
fine structure constant etc. The parameter $\lambda$ also depends on less fundamental parameters, including galaxy-dependent
quantities such as halo temperature.
Although in principle $\lambda$ can be a complicated function of these parameters, some simple arguments favour an
approximate scaling with the halo temperature, $\lambda \propto 1/\sqrt{T}$.
Such a simple scaling is roughly consistent with the analysis of the LITTLE THINGS dwarfs and THINGS spirals presented here.
Deviations from this scaling are possible and can in principle be calculated within a given dissipative particle model, and could
thereby provide an important means of testing and constraining the underlying particle physics.

\vskip 0.7cm
\noindent
{\bf Acknowledgments}

\vskip 0.1cm
\noindent
The author would like to thank: S. Oh and W. de Blok
for making available rotation curve data files. The author would
also like to thank S. Oh and also I. D. Karachentsev for some helpful correspondence.
This work was supported by the Australian Research Council.

\end{document}